\documentclass[
pre,
aps,
twocolumn,
english,
superscriptaddress,
floatfix,
amsmath,amssymb,
longbibliography,
nofootinbib
]{revtex4-2}

\usepackage{graphicx}
\usepackage{dcolumn}
\usepackage{bm}
\usepackage[hidelinks]{hyperref}
\usepackage{soul}
\hypersetup{colorlinks = true, allcolors = blue}
\usepackage{mathtools}
\usepackage{xcolor}
\newcommand{\sn}[1]{\textcolor{blue}{#1}}
\graphicspath{{./Figures/}}

\newcommand{\km}[1]{\textcolor{blue}{#1}}

\usepackage{xr-hyper}
\usepackage{lettrine}
\begin{document}

\preprint{APS/123-QED}

\title{Thickness effects in the electromechanical stability\\ of charged biological membranes}

\author{Sirui Ning}
\thanks{These authors contributed equally.}
\affiliation{Department of Chemical and Biomolecular Engineering, University of California, Berkeley, CA 94720, USA}

\author{Yannick A. D. Omar}
\thanks{These authors contributed equally.}
\affiliation{Department of Chemical Engineering, Massachusetts Institute of Technology, Cambridge, MA 02139, USA}

\author{Karthik Shekhar}
\email{kshekhar@berkeley.edu}
\affiliation{Department of Chemical and Biomolecular Engineering, University of California, Berkeley, CA 94720, USA}
\affiliation{Helen Wills Neuroscience Institute, California Institute for Quantitative Biosciences, QB3, Center for Computational Biology, University of California, Berkeley, CA 94720, USA}
\affiliation{Biological Systems Division, Lawrence Berkeley National Laboratory, Berkeley, CA 94720, USA}

\author{Kranthi K. Mandadapu}
\email{kranthi@berkeley.edu}
\affiliation{Department of Chemical and Biomolecular Engineering, University of California, Berkeley, CA 94720, USA}
\affiliation{Chemical Sciences Division, Lawrence Berkeley National Laboratory, CA 94720, USA}

\date{\today}

\begin{abstract}
\vspace{0.1in}
Understanding how electric fields destabilize biological membranes is important for electroporation-based technologies and bioelectronic interfaces. However, theoretical descriptions of this phenomenon remain fragmented. Existing theories treat either electrostatics in membranes of finite thickness or electrohydrodynamic flows at idealized zero-thickness interfaces, leaving unresolved a unified description that simultaneously incorporates finite membrane thickness, surface charge, and bulk electrohydrodynamics. Here, we apply a recently-developed, dimension-reduction framework that captures the coupled electrohydrodynamic and mechanical effects governing height fluctuations of a charged lipid bilayer of thickness $\delta$ in an electrolyte characterized by Debye screening length $\lambda$. We derive voltage- and charge-dependent renormalizations of the effective surface tension and bending rigidity, along with a dispersion relation governing undulatory instabilities. A wide range of prior theoretical results arise as limiting cases of our more general theory when finite-thickness effects are neglected or screening is asymptotically strong. The key new contribution arises from traction moments generated across the finite membrane thickness, which are absent in zero-thickness descriptions. Under physiological screening ($\delta/\lambda\sim 4$), these contributions account for more than $>70\%$ of the total electrostatic correction to both surface tension and bending rigidity. The theory further reveals that surface charges can stabilize the membrane at physiological ionic strengths, increasing the effective tension and shifting electroporation thresholds in a manner that depends on charge asymmetry between the leaflets.
\end{abstract}

\maketitle

\section{Introduction}
\label{sec:introduction}
\lettrine{T}{he} interaction of electric fields with biological membranes is a fundamental feature of cellular physiology, underlying processes such as action potential propagation, synaptic transmission, mechanosensation, and---under strong applied fields---electroporation~\cite{hille1992, bezanilla2008membrane, buzsaki2012origin, kotnik2019membrane}. Despite being only ${4 - 7 \text{ nm}}$ thick, biological membranes routinely sustain transmembrane voltages of ${50 - 200 \text{ mV}}$ in physiological conditions and up to nearly ${1.0 \text{ V}}$ under pulsed stimulation~\cite{Needham1995,hemmerle2016reduction}. These voltages generate electric fields that act on charges in the membrane and the surrounding electrolyte, producing electromechanical responses enabled by the membrane's unusual combination of in-plane fluidity~\cite{singer1972fluid} and out-of-plane elasticity~\cite{helfrich1973elastic}. 

Electromechanical behaviors of membranes have been probed across multiple experimental settings. Applied voltages can reduce the tension required for rupture, allowing bilayers to rupture at area dilations of about ${2 - 3 \%}$ (comparable to values observed in purely mechanical experiments) when the transmembrane potential approaches ${0.2 - 1.1 \, \text{V}}$~\cite{Needham1995, Riske2005,weaver1981decreased}. Under electric fields, giant unilamellar vesicles (GUVs) can deform and develop pores, with dynamics governed by field strength and ionic composition~\cite{Riske2005, Dimova2009, Aranda2008,aleksanyan2023assessing}. The presence of surface charges---either from anionic lipids such as POPG or from adsorbed ions---can further modulate bending rigidity, surface tension, and electroporation thresholds~\cite{Shoemaker2002_IntramembraneElectrostatics}. Thus, membranes act not just as passive capacitors but as responsive electromechanical materials in which electric fields, ionic reorganization, and mechanical stresses are intimately coupled.

From both theoretical and experimental perspectives, a central question is how applied electric fields modify surface tension $\Lambda$ and bending rigidity $k_{\rm b}$, the principal mechanical properties relevant for membrane shape fluctuations and stability. Addressing this requires a framework that self-consistently couples electrostatics, hydrodynamics, and mechanics under applied electric fields. Three principal classes of theoretical approaches have been developed to quantify electrostatic corrections to membrane mechanical properties, differing in how they treat the membrane and charge dynamics in the surrounding electrolyte: 

The \emph{first} approach derives electrostatic corrections under equilibrium conditions within the Helfrich membrane elasticity framework~\cite{helfrich1973elastic, ouyang2026beautymathematicshelfrichsbiomembrane, PhysRevA.39.5280} by solving the Poisson–Boltzmann equations near curved, charged interfaces and expanding the resulting electrostatic free energy in terms of curvature~\cite{winterhalter1988effect, winterhalter1992bending, lekkerkerker1989contribution}. This approach provides equilibrium corrections to $\Lambda$ and $k_{\rm b}$, but is intrinsically static and does not resolve membrane dynamics or bulk charge transport and hydrodynamics. 

The \emph{second} approach resolves the membrane as a dielectric slab of finite thickness and computes electrostatic corrections by solving Poisson’s equation across the membrane and surrounding electrolyte. By integrating three-dimensional Maxwell stresses and (in some models) explicitly enforcing a local mechanical stress balance, this formulation captures internal field gradients and inter-leaflet coupling more faithfully than zero-thickness models~\cite{ambjornsson2007applying, loubet2013electromechanics}. However, it remains restricted to static or quasi-static configurations and does not self-consistently incorporate membrane hydrodynamics, ionic fluxes, or the spatiotemporal evolution of double layers. 

The \emph{third} approach treats the bilayer as an effective zero-thickness interface coupled to electrolyte dynamics. In the limit of thin Debye layers, one may either adopt the classical Taylor–Melcher leaky dielectric framework for interfacial electrohydrodynamics~\cite{taylor1965stability, melcher1969electrohydrodynamics, sens2002undulation, schwalbe2011lipid, young2014long}, or use the Poisson–Nernst–Planck (PNP) equations, which explicitly retain diffuse charge layers and ionic transport~\cite{Ziebert2010_ZeroThickness,ziebert2010poisson, lacoste2009electrostatic, yu2025instabilityfluctuatingbiomimeticmembrane}. These models investigate dynamic electromechanical behaviors while idealizing the membrane as a surface of vanishing thickness, and therefore cannot consistently resolve internal dielectric structure or inter-leaflet charge asymmetry. 
Moreover, collapsing the bilayer into a single interface precludes a natural representation of distinct surface charge densities on the two leaflets, which is important for biological membranes, whose lipid compositions are inherently asymmetric.
Consequently, existing frameworks either resolve finite-thickness electrostatics under static conditions or model electrohydrodynamic effects at zero-thickness interfaces, but do not simultaneously capture thickness-dependent Maxwell stresses, dynamic ionic reorganization, and coupled hydrodynamics between the membrane and the bulk.

To address these limitations, we recently developed a self-consistent theoretical framework, termed the $(2+\delta)$-dimensional theory, that bridges full three-dimensional (3D) electromechanical descriptions and strict zero-thickness surface models~\cite{omar1,omar2, omar3}. The $(2+\delta)$-dimensional framework systematically reduces the 3D balance laws of mass, momentum, and electrostatics to an effective surface description by expanding all fields spectrally across the membrane thickness. This procedure retains finite-thickness contributions to intramembrane electric fields and stresses while enforcing continuity of traction and velocity, as well as the electrostatic boundary conditions at the membrane–electrolyte interfaces. The resulting dimensionally-reduced equations couple in-plane viscous flow, out-of-plane elasticity, and internal dielectric structure without introducing ad hoc boundary conditions. Moreover, because the two membrane interfaces are treated explicitly, the framework also allows independent leaflet charge densities to enter the electrostatic boundary conditions, enabling a consistent treatment of charge asymmetry across the bilayer. 

\begin{figure}
\centering
\includegraphics[width=0.8\linewidth]{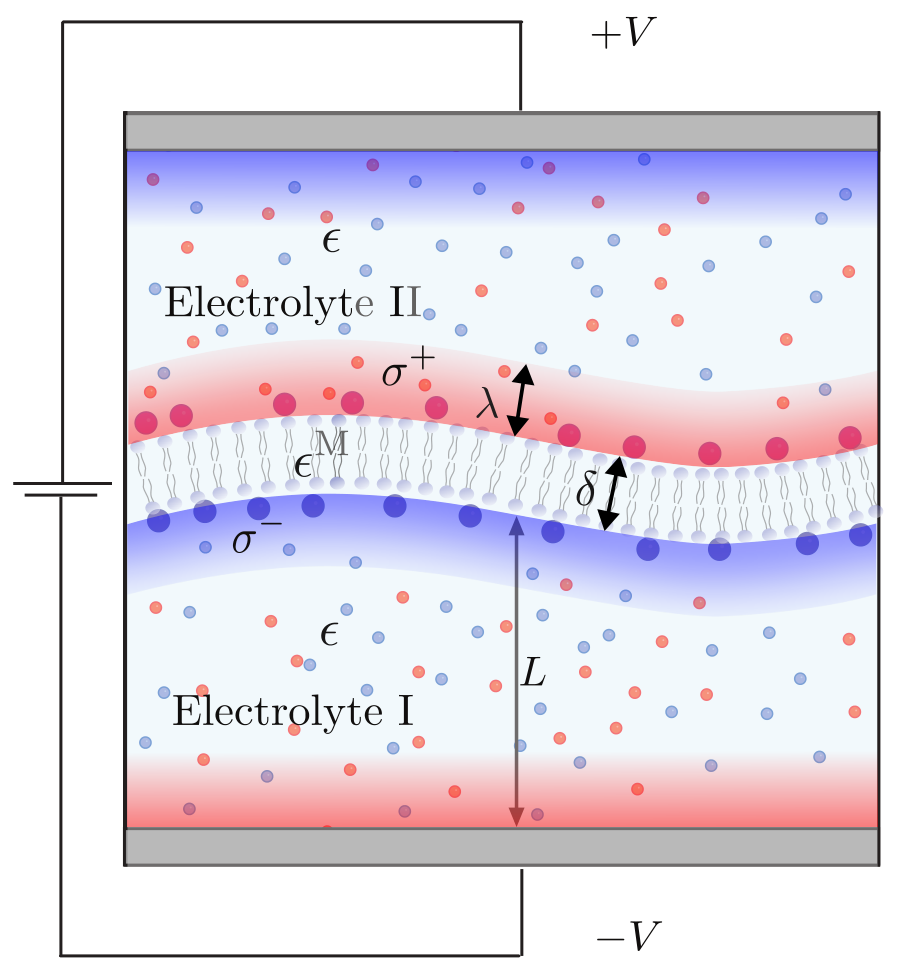}
\caption{A lipid membrane of thickness $\delta$ bearing intrinsic surface charges (large purple spheres on upper/lower leaflets) is subjected to a transverse voltage ($+V$ at the upper electrode, $-V$ at the lower electrode). Each electrode is placed at a distance ${L \gg \lambda}$ from the nearest membrane surface, when the membrane is flat. The upper and lower membrane leaflets carry surface charge densities $\breve{\sigma}^+$ and $\breve{\sigma}^-$, respectively. The applied field induces electrostatic interactions with the surrounding electrolyte ions (blue and orange spheres), modifying the membrane's intrinsic surface tension $\Lambda$ and bending rigidity $k_\text{b}$ to their effective values $\Lambda^{\mathrm{eff}}$ and $k_{\text{b}}^{\mathrm{eff}}$. These field-renormalized mechanical properties regulate membrane undulations $h(x,y,t)$ and stability.}
\label{fig:membrane}
\end{figure}

In this work, we apply the $(2+\delta)$-dimensional framework to analyze the electromechanical stability of a finite-thickness lipid membrane bearing asymmetric surface charges and subjected to an applied transmembrane voltage (Fig.~\ref{fig:membrane}). By analyzing the coupled electrostatic, hydrodynamic, and elastic equations, we derive renormalized expressions for the effective surface tension and bending rigidity and obtain closed-form dispersion relations governing membrane fluctuations and their transition to undulatory instabilities. We explicitly separate the contributions from applied voltage, fixed surface charge, and their coupling, and show that earlier electrostatic and electrohydrodynamic models~\cite{lacoste2009electrostatic, winterhalter1988effect, Ziebert2010_ZeroThickness, ambjornsson2007applying, loubet2013electromechanics, yu2025instabilityfluctuatingbiomimeticmembrane} are recovered as limiting cases when thickness-dependent stresses are neglected or when electric-field screening becomes asymptotically strong. A central finding is that in regimes relevent for physiology or biological experiments where $\delta/\lambda=O(1)$, the thickness-dependent contributions account for the majority of the electrostatic renormalization of both $\Lambda$ and $k_{\rm b}$, thereby determining the onset and quantitative boundary of electromechanical instabilities in biological membranes. 

\section{Theory}
\label{sec:theory}

\subsection{Problem setup}
Consider an initially flat biological membrane of thickness $\delta$ and permittivity $\varepsilon_{\mathcal M}$ separating identical, 1:1 electrolytes (permittivity $\varepsilon_{\mathcal B}$, viscosity $\mu$) above/below. The electrolytes, each with bulk concentration $C_0$, screen electric fields with Debye length ${\lambda=(\varepsilon_{\mathcal B}k_{\rm B}T/2e^2C_0)^{1/2}}$. Under physiological conditions, ${\lambda \sim 1 \text{ nm}}$, although vesicle electromechanics experiments routinely access regimes with $\lambda$ spanning $\sim$ ${0.5-10 \text{ nm}}$ through controlled variation of salt concentrations~\cite{dimova2019giant}.  A potential difference of $2V$ is imposed between distant electrodes ($+V$ above, $-V$ below). Throughout this work we assume the electrodes are asymptotically far from the membrane (${L \gg \lambda}$), so that potential drops across the electrode Debye layers are negligible and the electrolyte can be treated as effectively unbounded.

The membrane has intrinsic (base) surface tension $\Lambda$ and bending modulus $k_{\rm b}$. The upper and lower leaflets carry spatially uniform surface charge densities $\breve{\sigma}^\pm$ when the membrane is flat. The different permittivities of the membrane and bulk lead to a dielectric mismatch, quantified by the dimensionless parameter ${\Gamma \coloneqq \varepsilon_{\mathcal B} / \varepsilon_{\mathcal M} \sim 20-40}$. Table~\ref{tab:params} summarizes the physical parameters featured in this work and their typical physiological ranges. Our goal is to determine, in the regime of small fluctuations, how electrostatic interactions modify the \emph{effective} surface tension $\Lambda^{\mathrm{eff}}$ and bending modulus $k_{\rm b}^{\mathrm{eff}}$, and to identify conditions under which modifications in these parameters lead to electromechanical instabilities.
\begin{table}[h]
\caption{\label{tab:params}%
List of parameters, symbols, and typical values used in this work. Typical parameter values are taken from standard membrane biophysics reviews and textbooks~\cite{helfrich1973elastic, seifert1997configurations, dimova2019giant, milo2015cell, Holmes_2011, omar2, omar3, sahu2022irreversible, TakatoriSahu2020}
}
\begin{ruledtabular}
\resizebox{0.9\linewidth}{!}{%
\begin{tabular}{lcc}
    Parameter & Symbol & Typical Range \\
    \hline
    Membrane thickness & $\delta$ & $4\ \text{nm}$ \\
    Fluid shear viscosity & $\mu$ & $10^{-3}\ \text{pN}\cdot\mu\text{s}/\text{nm}^2$  \\
    Fluid bulk viscosity & $w$ & $2.5\cdot10^{-3}\ \text{pN}\cdot\mu\text{s}/\text{nm}^2$  \\
    Membrane shear viscosity & $\zeta=\delta \mu$ & $4\cdot 10^{-3}\ \text{pN}\cdot\mu\text{s}/\text{nm}$ \\
    Membrane bulk viscosity & $\bar{w}=\delta w$ & $10^{-2}\ \text{pN}\cdot\mu\text{s}/\text{nm}$ \\
    Compression modulus & $\bar{k}_\mathrm{c}=\frac{2k_\mathrm{b}}{\delta^2}$ & 10.5 pN/nm \\
    Membrane density & $\rho_{0}$ & $10^{-9}\ \text{pg}/\text{nm}^{3}$ \\
    Membrane areal density & $\breve{\rho}_s=\delta\rho_0$ & $4\cdot 10^{-9}\ \text{pg}/\text{nm}^{2}$ \\
    Dielectric mismatch & $\Gamma \coloneqq \varepsilon_{\mathcal{B}} /\varepsilon_{\mathcal{M}}$ & 20 \\
    Bending modulus & $k_{\text{b}}$ & $84\ \text{pN}\cdot\text{nm}$ \\
    Membrane surface tension & $\Lambda$ & $4\times 10^{-3}\ \text{pN}/\text{nm}$\sn{\footnote{This value lies at the lower end of typical membrane tensions; lipid membrane tensions can reach values as high as $\sim 1\ \text{pN}/\text{nm}$~\cite{doi:10.1073/pnas.2221541120}.}}  \\
    Membrane permittivity & $\varepsilon_{\mathcal M}$ & $1.4\times 10^{-3}\ \text{e}^{2}/(\text{pN}\cdot\text{nm}^{2})$ \\
    Salt concentration & $C_0$ & $1 - 150$ mM \\
    Debye length & $\lambda$ & $1 - 10$ nm \\
    Surface charge density & $\breve{\sigma}^\pm$ & $0.06 - 0.245$ e/nm$^2$\footnote{This range corresponds to an estimated charged-lipid number fraction of $\sim 5 - 20 \%$ on each leaflet, computed as the fraction of charged lipids among all surface lipids assuming an average molecular area of 80~$\text{\AA}^2$ per lipid~\cite{10.7554/eLife.04366}.} \\
    Thermal voltage & $\phi_{\rm T}$ & 25 mV \\
\end{tabular}%
}
\end{ruledtabular}
\label{table1}
\end{table}

\subsection{Bulk ion transport and electrohydrodynamics}
The bulk fluids above ($+$) and below ($-$) the membrane are incompressible Newtonian fluids, each containing a dilute, symmetric, monovalent electrolyte.  Ionic concentrations $c_i,\, i\in\{1,2\}$ evolve according to the Poisson-Nernst–Planck (PNP) equations~\cite{Nernst1888, Nernst1889, Planck1890, fong2020transport}:
\begin{subequations}
\label{eq:PNP}
\begin{align}
    \frac{\mathrm{D}c_i^{\pm}}{\mathrm{D}t}+c_i^{\pm}\nabla\!\cdot\!\boldsymbol{u}^{\pm}
&= -\nabla\!\cdot\!\boldsymbol{J}_i^{\pm}\ , \label{eq:PNP1}\\
    \boldsymbol{J}_i^{\pm}
&=-D\!\left(\nabla c_i^{\pm}+\frac{z_i \mathrm{e}}{k_{\rm B}T}c_i^{\pm}\nabla\phi_{\mathcal{B}}^{\pm}\right) \label{eq:PNP2} \ , \\
\Delta\phi^\pm_{\mathcal{B}} &= -q_{\!f}^{\pm} / \varepsilon_{\mathcal{B}} \ ,\label{eq:PNP3}
\end{align}
\end{subequations}
where $\boldsymbol{u}^\pm$ are bulk fluid velocities, $\boldsymbol{J}_i^{\pm},\,i\in\{1,2\}$ are the mass fluxes, $D$ is the ionic diffusivity (assumed equal for both species), $k_{\rm B} T$ is the thermal energy scale, $\mathrm{e}$ is the fundamental charge, $z_i$ are the valencies with ${z_1=1}$ and ${z_2=-1}$, and $q_{f}^{\pm}$ are the free charge densities in the two bulk domains. Note that ${\mathrm{D}/\mathrm{D}t \coloneqq \partial_t+\boldsymbol{u}^{\pm}\!\cdot\!\nabla}$ is the material derivative. Equation~\eqref{eq:PNP1} expresses the conservation of mass for species $i$, Eq.~\eqref{eq:PNP2} describes the ionic flux as the sum of contributions from Fickian diffusion and electromigration, and Eq.~\eqref{eq:PNP3} is Poisson's equation relating the electric potential to the free charge density in the bulk ${q_{\!f}^\pm = \mathrm{e}(c_1^{\pm}-c_2^{\pm})}$.

At low Reynolds number, the bulk fluid motion is described by the incompressible Stokes equations~\cite{kovetz2000electromagnetic, fong2020transport, omar2}:
\begin{equation}
\nabla \cdot \boldsymbol{u}^{\pm} = 0, \quad \nabla \cdot \boldsymbol{\sigma}_{\mathcal{B}}^\pm = \boldsymbol{0} \ ,
\label{eq:Stokes_bulk}
\end{equation}
where the total bulk stress tensor, $\boldsymbol{\sigma}_{\mathcal{B}}^\pm$, combines hydrodynamic and electrostatic (Maxwell) contributions:
\begin{align}
\boldsymbol{\sigma}_{\mathcal{B}}^\pm\! &=\underbrace{\!-p^{\pm}\boldsymbol{I} + \mu\left(\nabla \boldsymbol{u}^{\pm} + (\nabla \boldsymbol{u}^{\pm})^{T}\right)}_{\text{Hydrodynamic stress}} \\ & \qquad + \; \underbrace{\varepsilon_{\mathcal{B}}\left(\boldsymbol{E}_{\mathcal{B}}^\pm \otimes \boldsymbol{E}_{\mathcal{B}}^\pm - \frac{1}{2} |\boldsymbol{E}_{\mathcal{B}}^\pm |^2 \boldsymbol{I}\right)}_{\text{Maxwell stress}} \ . \label{eq:sigmaB}
\end{align}
The first term represents pressure and viscous stress in the fluid, while the second captures the stresses induced by electric fields. The interplay of these stresses governs the electrohydrodynamic traction transmitted to the membrane surfaces. In Eq.~\eqref{eq:sigmaB}, $p^{\pm}$ is the pressure, $\mu$ is the shear viscosity of the fluid, and ${\boldsymbol{E}_{\mathcal{B}}^\pm=-\nabla\phi_{\mathcal{B}}^\pm}$ is the bulk electric field. Taken together, Eqs.~\eqref{eq:PNP1}--\eqref{eq:sigmaB} define the bulk mass, charge, and momentum transport equations that couple self-consistently to the membrane through the $(2+\delta)$ theory.

\subsection{Overview of the ${(2+\delta)}$-dimensional framework}
The ${(2+\delta)}$-dimensional theory was derived in Refs.~\cite{omar1,omar2, omar3}, and only the essential physical ideas are summarized here. The membrane is modeled as a thin film of thickness $\delta$ and permittivity $\varepsilon_{\mathcal M}$ that behaves as a two-dimensional viscous fluid in-plane and as an elastic shell out-of-plane~\cite{omar1, omar2, omar3} (Fig.~\ref{fig:membrane_delta}). Its geometry is described by its midsurface parameterized by curvilinear coordinates ${\{\theta^\alpha \}}$, with tangent basis $\{\mathbf{a}_\alpha \}$, unit normal $\mathbf{n}$; throughout, Greek indices $(\alpha,\beta,\gamma,\ldots)$ span the in-plane coordinates $1,2$. Standard quantities from differential geometry are used to characterize the surface: the metric tensor $a_{\alpha \beta}$, curvature tensor $b_{\alpha \beta}$, mean curvature $H$, and Gaussian curvature $K$.

We assume that material lines initially normal to the midsurface remain normal and unstretched during deformation~\cite{timoshenko1959theory}. Thus, any material point can be described by ${\boldsymbol{x}(\theta^\alpha,\theta^3,t)=\boldsymbol{x}_0(\theta^\alpha,t)+\theta^3\mathbf n}$, with ${\theta^3 \! \in[-\delta/2,\delta/2]}$. In the ${(2+\delta)}$-dimensional theory, fields are expanded spectrally in the normalized thickness coordinate ${\Theta \coloneqq (2/\delta)\theta^3\in[-1,1]}$ using Chebyshev polynomials $P_k(\Theta)$. For example,
\begin{equation}
\boldsymbol{x} \approx \sum_{k=0}^{m} \boldsymbol{x}_k(\theta^{\alpha}, t) P_k[\Theta] \ , \,\, \boldsymbol{v} \approx \sum_{k=0}^{m} \boldsymbol{v}_k(\theta^{\alpha}, t) P_k[\Theta]  \ ,
\label{eq:twoplusdelta_rep}
\end{equation}
where $\boldsymbol{x}$ and $\boldsymbol{v}$ are position and velocity fields. Here, ${P_k(\cdot)}$ are the Chebyshev polynomials of the first kind~\cite{boyd2001chebyshev}. 

The reduction in dimensionality arises because all dependence on the thickness coordinate is captured by the prescribed basis functions $P_k$, while the coefficients $\boldsymbol{x}_k$ or $\boldsymbol{v}_k$ depend only on midsurface coordinates and time. For sufficiently thin membranes, fields vary weakly across the thickness, so the expansions~\eqref{eq:twoplusdelta_rep} can be truncated at low order of $m$. Overall, this formulation is valid when the membrane behaves as a geometrically thin shell, i.e. $(\delta H)^2 \ll 1$ and $\delta^2|K| \ll 1$, and that the characteristic in-plane variation length scales of curvature, stress, and electric potential are large compared to $\delta$. Under these conditions, the theory reduces the full 3D electromechanical problem to an effective 2D surface description while retaining explicit thickness-dependent corrections. This allows finite-thickness effects modulated by transmembrane potential differences, dielectric mismatch, and Maxwell stresses, to be incorporated self-consistently into the membrane mechanical equations.

\subsection{$(2+\delta)$-dimensional equations for membrane electrostatics}
Inside the membrane, which is assumed to contain no free charge, the potential satisfies Poisson’s equation $\varepsilon_{\mathcal M}\nabla^2\phi_{\mathcal M}=0$. The interfacial conditions at the upper and lower membrane surfaces $\mathcal{S}^{\pm}$ enforce continuity of the potential and the electric displacement (Gauss's law). Because the membrane is isolating, a no-flux boundary condition is imposed as well:
\begin{subequations} \label{membrane_bc}
\begin{align}
\left. \left( \phi_{\mathcal{B}}^\pm - \phi_{\mathcal{M}} \right) \right|_{{\mathcal{S}^\pm}} &= 0 \ , \label{membrane_phi_continuity}\\
 \left. \boldsymbol{n}^\pm\cdot (\varepsilon_\mathcal{B}\boldsymbol{E}^\pm_\mathcal{B}-\varepsilon_\mathcal{M}\boldsymbol{E}_\mathcal{M}) \right|_{\mathcal{S}^\pm} &= \sigma^\pm \ ,\\
 \left. \boldsymbol{n}^\pm\cdot\boldsymbol{J}_i^{\pm}\right|_{\mathcal{S}^\pm}&=0\ ,
 \label{membrane_jump_condition}
 \end{align}
\end{subequations} 
where ${\boldsymbol{n}^\pm=\pm \boldsymbol{n}}$ are the surface normals, and ${\boldsymbol{E}_{\mathcal{M}}=-\nabla\phi_{\mathcal{M}}}$ is the electric field inside the membrane. Retaining the two interfaces $\mathcal S^\pm$ allows distinct leaflet charge densities $\sigma^\pm$ to appear directly in Gauss's law, which is not possible in zero-thickness approaches. Following Eq.~\eqref{eq:twoplusdelta_rep}, we introduce a low-order expansion for $\phi_{\mathcal{M}}$:
\begin{equation}
\phi_{\mathcal{M}}(\theta^\alpha,\theta^3, t) = \sum_{k=0}^2 \phi_k(\theta^\alpha, t) P_k[\Theta(\theta^3)] \ .
\label{eq:phiM_twoplusdelta}
\end{equation}
In the $(2+\delta)$ theory, we retain modes up to ${m=2}$ for the electric potential, while truncating at ${m=1}$ for $\boldsymbol{v}$ and $\boldsymbol{x}$. For $\phi_\mathcal{M}$, keeping terms through $m=2$ is essential both to satisfy the two independent boundary conditions at $\mathcal{S}^\pm$ and to capture the $O(\delta)$ thickness moments of the electric field that enter the membrane Maxwell stress. Substituting this expansion into the membrane Poisson equation, applying Chebyshev orthogonality, and enforcing the boundary conditions~\eqref{membrane_bc} yields coupled surface equations for the coefficients $\phi_k$. 

Physically, this reduction transforms the three-dimensional electrostatic problem into an effective surface description in which the potential is represented by its leading coefficients: $\phi_0$ represents the baseline, thickness-independent component of the membrane potential, $\phi_1$ captures the transmembrane voltage drop, and $\phi_2$ accounts for quadratic variations associated with dielectric mismatch, geometric effects, and in-plane gradients. The boundary conditions~\eqref{membrane_bc} are enforced through the relations among these coefficients, ensuring the satisfaction of potential continuity and Gauss's law at each interface. The full equations are given in Ref.~\cite{omar2} and {summarized in Sec. II in the Supplemental Material (SM).}

\begin{figure}
    \centering
    \includegraphics[width=1\linewidth]{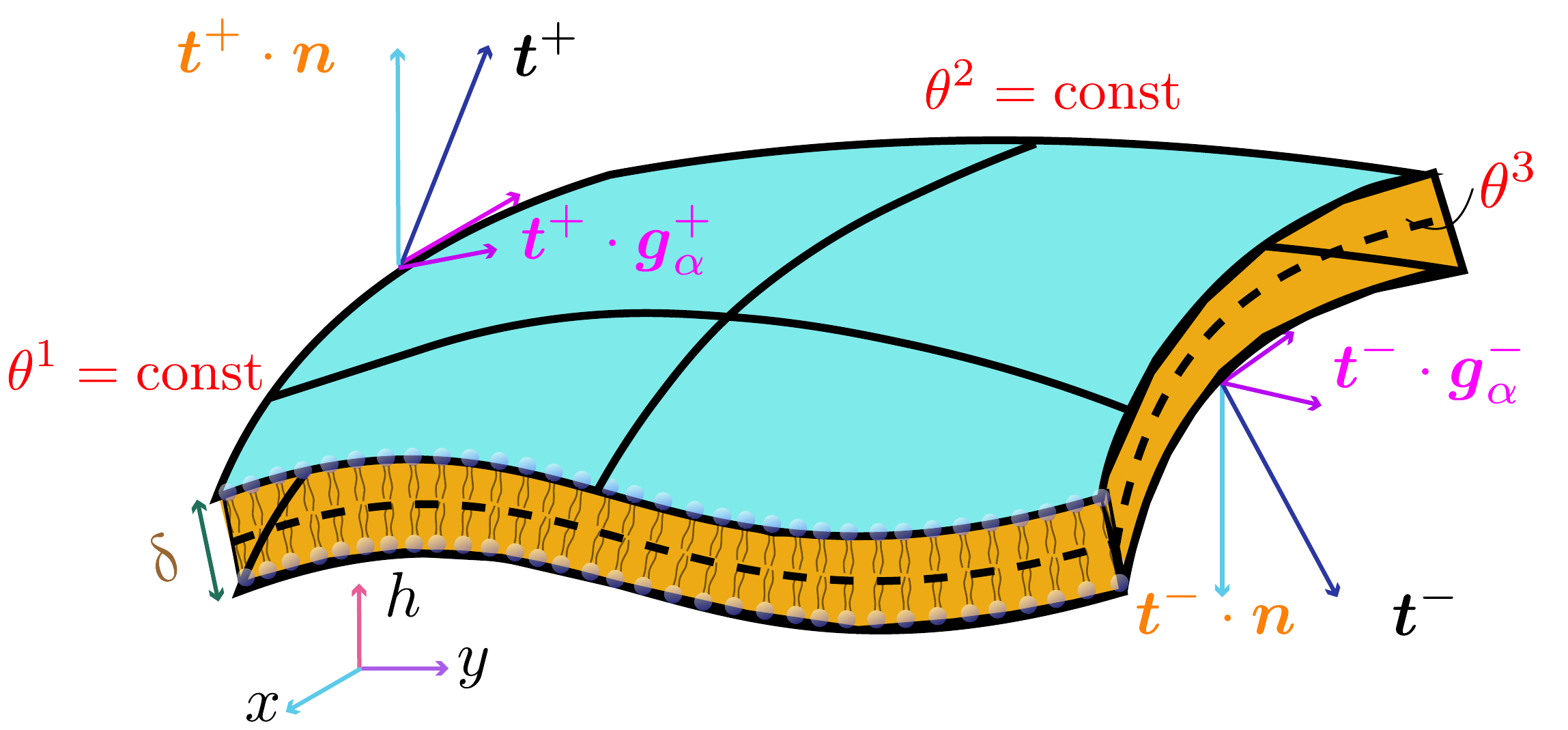}
    \caption{
Schematic of the membrane surface showing the local tangent basis vectors $\boldsymbol{g}^\pm_{\alpha}$ 
and the unit normal vector $\boldsymbol{n}$. 
The total traction vectors acting on the upper and lower membrane leaflets are 
denoted by $\boldsymbol{t}^{+}$ and $\boldsymbol{t}^{-}$, respectively. 
Their projections along the tangent directions 
$\boldsymbol{t}^{\pm}\!\cdot \boldsymbol{g}^\pm_{\alpha}$ and the normal direction 
$\boldsymbol{t}^{\pm}\!\cdot \boldsymbol{n}$ are indicated.
}
\label{fig:membrane_delta}
\end{figure}

Solving for $\phi_{\mathcal{M}}$ yields the electric field $\boldsymbol{E}_{\mathcal{M}}$, which defines the Maxwell stress tensor inside the membrane:
\begin{equation}
\begin{aligned}
\boldsymbol{\sigma}_{\mathcal{M}} &= \varepsilon_{\mathcal{M}}\left(\boldsymbol{E}_{\mathcal{M}}\otimes\boldsymbol{E}_{\mathcal{M}} - \tfrac{1}{2}|\boldsymbol{E}_{\mathcal{M}}|^2\boldsymbol{I}\right) \ .
\end{aligned}
\label{eq:sigmaM}
\end{equation}
At the upper and lower surfaces $\mathcal{S}^\pm$, the total traction vectors $\boldsymbol{t}^\pm$  (Fig.~\ref{fig:membrane_delta}) are defined as:
\begin{equation}
\boldsymbol{t}^\pm
= \left.(\boldsymbol{\sigma}_{\mathcal{B}}^\pm - \boldsymbol{\sigma}_{\mathcal{M}})\cdot\boldsymbol{n}^\pm\right|_{\mathcal{S}^\pm},
\label{traction_definition}
\end{equation}
where the bulk stress tensors $\boldsymbol{\sigma}_{\mathcal{B}}^\pm$ are defined in Eq.~\eqref{eq:sigmaB}, and include both hydrodynamic and Maxwell contributions. These tractions enter the membrane’s mechanical balance equations as described in the next section.
\subsection{$(2+\delta)$-dimensional equations of motion for the membrane}
The ${(2+\delta)}$ framework provides a unified description of membrane electromechanics by reducing the full three-dimensional balance laws to an effective two-dimensional surface formulation that retains finite-thickness corrections. Within this approach, the membrane is treated as a viscous-elastic surface coupled to the surrounding electrolytes through mechanical and electrostatic boundary conditions that determine the interfacial tractions $\boldsymbol{t}^\pm$ defined in Eq.~\eqref{traction_definition}.

In the inertialess regime, the governing equations follow from the balance of mass, in-plane momentum, and out-of-plane momentum. Mass conservation enforces area incompressibility through the continuity equation for the surface velocity, ${v^{\alpha}_{0;\alpha}-2vH =0}$, where $v^{\alpha}_0$ denotes the in-plane fluid velocity, $v$ is the normal velocity, and the subscript ``$;\alpha$'' indicates the covariant derivative. The in-plane momentum balance describes intramembrane tangential flows:
\begin{equation}\label{eq:tangential_balance}
\begin{aligned}
0 &= \Lambda_{,\beta}a^{\alpha\beta}
   + \frac{k_{\rm b}}{2}K_{,\beta}a^{\alpha\beta}
   + \pi_{;\beta}^{\beta\alpha}
   + (\boldsymbol{t}^+ + \boldsymbol{t}^-)\cdot\boldsymbol{a}^{\alpha} \\
  &\quad
   - \frac{\delta}{2}
     \left(2H\delta_{\gamma}^{\alpha}+b_{\gamma}^{\alpha}\right)
     (\boldsymbol{t}^+ - \boldsymbol{t}^-)\cdot\boldsymbol{a}^{\gamma} \ ,
\end{aligned}
\end{equation}
where $\pi^{\alpha\beta}$ are the membrane viscous stress tensor components. The out-of-plane momentum balance, known as the \emph{shape equation}, governs the normal force balance on the midsurface and controls membrane deformation and stability:
\begin{equation}\label{eq:shape_equation}
\begin{aligned}
0 &= 2\Lambda H 
     - k_{\rm b}(4H^2 - 3K)H 
     - k_{\rm b}\Delta_s H 
     + \pi^{\alpha\beta}b_{\alpha\beta} \\
  &\quad
     + Q^{\alpha}_{;\alpha}
     + (\boldsymbol{t}^+ + \boldsymbol{t}^-)\!\cdot\!\boldsymbol{n}
     - \delta H(\boldsymbol{t}^+ - \boldsymbol{t}^-)\!\cdot\!\boldsymbol{n} \ .
\end{aligned}
\end{equation}
In Eq.~\eqref{eq:shape_equation}, the first three terms represent elastic restoring forces from surface tension and bending rigidity. The fourth term $\pi^{\alpha\beta}b_{\alpha\beta}$ couples in-plane viscous stresses to curvature. The fifth term involves the divergence of $Q^{\alpha}$, which is the transverse shear force resultant, obtained by integrating across the membrane thickness the deviation of the membrane stress from the bulk electrolyte stress and is given by:
\begin{equation}\label{eq:def_Q_alpha}
Q^{\alpha} = \delta \left(\frac{1}{2}(\boldsymbol{t}^+ - \boldsymbol{t}^-) - \frac{\delta H}{4}\left(\boldsymbol{t}^+ + \boldsymbol{t}^-\right) \right) \cdot \boldsymbol{a}^{\alpha} \ .
\end{equation}
The sixth and seventh terms involve normal components of interfacial tractions; notably, the ${(2+\delta)}$-dimensional theory introduces an ${O(\delta)}$ correction proportional to the traction difference across the membrane, which vanishes in the zero-thickness limit. 

Equations~\eqref{eq:tangential_balance} and ~\eqref{eq:shape_equation} are the central dynamical relations of the $(2+\delta)$-dimensional theory, representing tangential and normal force balances that incorporate electrohydrodynamic coupling and traction asymmetry in a self-consistent manner.

\subsection{Analysis of normal modes}
To analyze small fluctuations, we adopt the Monge representation (Fig.~\ref{fig:membrane_delta}) and express the membrane midsurface by ${\boldsymbol{x}_0(x,y,t) = x \boldsymbol{e_x} + y \boldsymbol{e_y} + \epsilon h(x,y,t) \boldsymbol{e_z}}$, where ${\epsilon \ll 1}$ characterizes the small amplitude of the deformation. We then linearize all fields about the flat equilibrium state ${\breve{h}=0}$. Linearizing Eqs.~\eqref{eq:tangential_balance} and~\eqref{eq:shape_equation} to $O(\epsilon)$ yields the governing equations for membrane fluctuations. Full details are presented in Appendix A of the SM. To solve the linearized equations, we then decompose the height field into its normal modes, ${h(x,y,t)=\sum_{\boldsymbol{q}}h_{\boldsymbol{q}} e^{\omega t+ i\boldsymbol{q}\cdot \boldsymbol{x}}}$, where $\boldsymbol{q}$ is the in-plane wavevector with magnitude ${q= | \boldsymbol{q} |}$. This decomposition diagonalizes the problem, reducing the dynamics at each $\boldsymbol{q}$ to an eigenvalue problem for the growth rate $\omega(q)$.

\begin{figure*}[t!]
\centering
\includegraphics[width=0.9\linewidth]{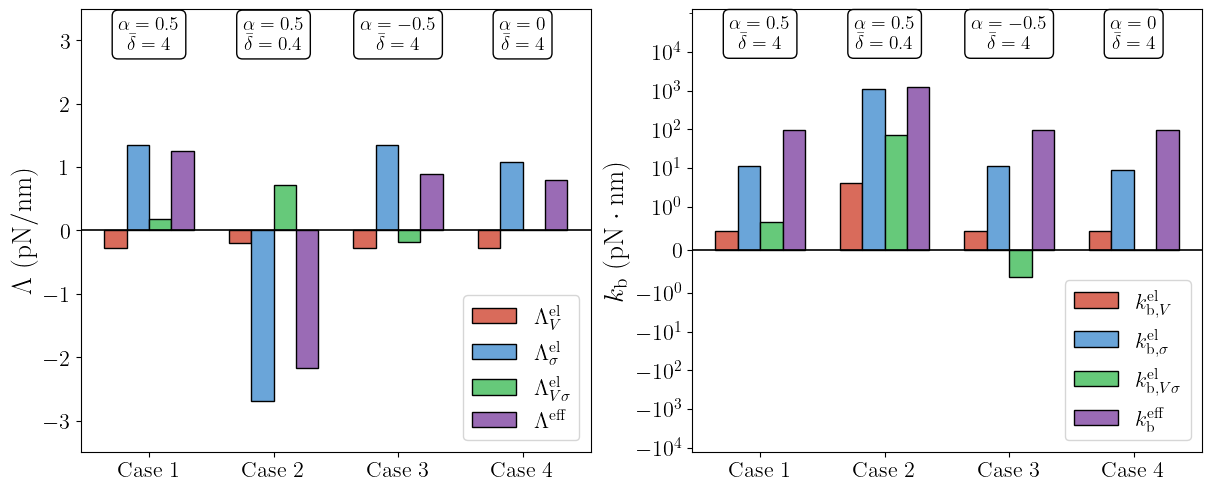}
\caption{
Bar charts decomposing the effective surface tension (left) and bending rigidity (right) into their voltage-driven ($V$), surface-charge-driven ($\sigma$), and mixed (${V\sigma}$) contributions for four representative membrane conditions (Cases~1--4). The dimensionless voltage is fixed at $\bar{V}=3$, and the mean dimensionless surface charge density is ${\langle \bar{\sigma}\rangle=-3.6}$. The corresponding charge asymmetry $\alpha$ and dimensionless thickness $\bar{\delta}$ for each case are indicated above the bars. For the surface tension, the voltage contribution is destabilizing in all cases. The surface-charge contribution is stabilizing when ${\bar{\delta}>1}$ and destabilizing when ${\bar{\delta}<1}$. Charge asymmetry further modulates stability: ${\alpha>0}$ (outer leaflet more negatively charged) stabilizes the membrane, whereas ${\alpha<0}$ (inner leaflet more negatively charged) promotes destabilization for ${\bar{V}>0}$.
The bending rigidity panel is shown on a symmetric logarithmic scale to resolve contributions spanning multiple orders of magnitude.
}
\label{fig:barplot}
\end{figure*}

\section{Results}
\label{sec:results}

\subsection{The dispersion relations}

Solving the coupled system of equations for $\omega(q)$ yields the dispersion relations (see SM Sec. III for details):
\begin{equation}
    \omega(q)=-\frac{q}{\mu(4+q^2\delta^2)}(\Lambda^{\text{eff}}+\frac{k^{\text{eff}}_{\text{b}}}{2}q^2) \ ,\label{dispersion}
\end{equation}
where $\Lambda^{\rm eff}$ and $k_{\rm b}^{\rm eff}$ are the voltage- and charge-renormalized tension and bending modulus, respectively. In the absence of electrostatic interactions, ${\Lambda^{\rm eff}=\Lambda>0}$ and ${k_{\rm b}^{\rm eff}=k_{\rm b}>0}$, so ${\omega(q)<0}$ for all $q$, and fluctuations decay monotonically. Electrostatic stresses modify these effective moduli and can drive $\Lambda^{\rm eff}$ negative, leading to long-wavelength undulation instabilities. The full expressions for $\Lambda^{\rm eff}$ and $k_{\rm b}^{\rm eff}$ are presented in Sec. III of the SM. It must be noted that the term $q^2\delta^2$ in the denominator of Eq.~\eqref{dispersion} is associated with additional dissipative effects arising from capturing the membrane thickness, recently described in Ref.~\cite{lipel2025finite}. 

To organize parameters, we introduce the dimensionless quantities:
\begin{equation}
\begin{aligned}
    &\bar{V}=\frac{V}{\phi_{\text{T}}}, \quad \bar{\sigma}^{\pm}=\frac{\breve{\sigma}^\pm \delta}{\Gamma\varepsilon_\mathcal{M}\phi_{\text{T}}}, \quad \langle\bar{\sigma}\rangle=\frac{1}{2}(\bar{\sigma}^++\bar{\sigma}^-) \\
    & \bar{\delta} = \frac{\delta}{\lambda}, \quad \Delta \bar{\sigma}=\frac{1}{2}(\bar{\sigma}^+-\bar{\sigma}^-), \quad \alpha=\frac{\Delta\bar{\sigma}}{\langle\bar{\sigma}\rangle},
\end{aligned}
\end{equation}
where ${\phi_{\rm T}\approx 25\text{ mV}}$ is the thermal voltage, and $\alpha$ quantifies the charge asymmetry between leaflets. The characteristic surface charge density, ${{\Gamma\varepsilon_\mathcal{M}\phi_{\text{T}}/\delta} \approx 0.027 \text{ e/nm}^2 }$, corresponding to $\sim$2 \% of lipids in a leaflet carrying a single elementary charge.

In what follows, we analyze the renormalized moduli and therefore the stability behavior as functions of the dimensionless voltage $\bar{V}$, the charge asymmetry $\alpha$, and the mean surface charge density $\langle \bar{\sigma} \rangle$, while fixing the remaining material parameters ($\phi_{\rm T}$, $\delta$, $\varepsilon_\mathcal{M}$, $\Gamma$) at the values listed in Table~\ref{table1}. To elucidate the role of electrostatic screening, we consider characteristic Debye lengths $\lambda = 1\,\mathrm{nm}$ (strong screening with ionic concentration $C_0\sim 150 \, \rm{ mM}$) and $\lambda = 5 - 10\,\mathrm{nm}$ (weak screening with ionic concentration $C_0\sim 1 \, \rm{ mM}$), which enter the theory through the dimensionless thickness $\bar{\delta} = \delta/\lambda$.

\subsection{Applied voltage in the absence of surface charge}
Applying a voltage to a membrane with no fixed surface charge (${\langle\bar{\sigma}\rangle = \Delta \bar{\sigma} = 0}$) results in:
\begin{equation}
\begin{aligned}
\Lambda^{\text{eff}} &= \Lambda \underbrace{- \frac{2 \bar{V}^2 \phi_{\rm T}^2 \Gamma\varepsilon_\mathcal{M} ((-2 + 3\Gamma) \bar{\delta} + 2)}{\lambda(\Gamma \bar{\delta} + 2)^2}}_{\Lambda^{\rm el}_V}, \\
\\
k^{\text{eff}}_{\text{b}} &= k_{\text{b}} \underbrace{+ \frac{\bar{V}^2 \phi_{\rm T}^2 \Gamma \varepsilon_\mathcal{M} \lambda (\Gamma \bar{\delta}^3 + 8\bar{ \delta}^2 + 16\bar{\delta} + 12)}{2 (\Gamma \bar{\delta} + 2)^2}}_{k_{{\rm b},V}^{\rm el}} \ ,
\label{eq:correctionsV}
\end{aligned}
\end{equation}
where $\Lambda^{\rm el}_V$ and $k_{{\rm b},V}^{\rm el}$ denote corrections due to applied voltage alone. Accordingly, an external voltage generically \emph{reduces} the effective tension and \emph{increases} the bending modulus. Figure~\ref{fig:barplot} highlights these effects in four cases spanning different screening regimes and surface charge asymmetry. For example, at ${\bar{V}=3}, {\bar{\delta}=4}$, and base tension ${\Lambda=4\times10^{-3}}$ pN/nm (Table~\ref{table1}), one finds ${\frac{\Lambda^{\text{eff}}-\Lambda}{\Lambda}\approx -70}$, indicating a 7,000 \% decrease in surface tension. On the other hand, the increase in $k_{\rm b}$ is modest ($\sim 0.5$\%). Consequently, voltage-driven instabilities, when present, are tension-dominated and emerge at long wavelengths.

We compare our expressions (Eq.~\eqref{eq:correctionsV}) to the results of a recent study~\cite{yu2025instabilityfluctuatingbiomimeticmembrane}, which employs a zero-thickness electrokinetic framework.  To enable a direct comparison, we rewrite Eq.~(43) of Ref.~\cite{yu2025instabilityfluctuatingbiomimeticmembrane} in the notation and dimensional conventions adopted in the present work. Their result is:
\begin{equation}
\begin{aligned}
\Lambda^{\text{eff}} &= \Lambda - \frac{4 \bar{V}^2 \phi_{\rm T}^2 \Gamma\varepsilon_\mathcal{M} (\Gamma\bar{\delta}+1)}{\lambda (\Gamma \bar{\delta} + 2)^2} \ , \\
\\
k^{\text{eff}}_{\text{b}} &= k_{\text{b}} + \frac{2\bar{V}^2 \phi_{\rm T}^2 \Gamma \varepsilon_\mathcal{M} \lambda (\Gamma \bar{\delta}^3 + 6 \bar{\delta} + 9)}{3 (\Gamma \bar{\delta} + 2)^2} \ .
\label{eq:correctionsV_yu}
\end{aligned}
\end{equation}
Comparing Eqs.~\eqref{eq:correctionsV} and~\eqref{eq:correctionsV_yu} shows that while both theories predict the same signs for the corrections, the magnitudes differ: in the limit ${\bar{\delta} \geq 1}$ and $\Gamma \gg 1$, the tension corrections differ by a factor of $3/2$ and the bending corrections by $3/4$. This discrepancy arises from the $Q^\alpha_{;\alpha}$ term in Eq.~\eqref{eq:shape_equation}, which captures intra-membrane gradients in the transverse direction that zero-thickness theories neglect. The quantitative nature of the differences, which can be substantial under physiological conditions, are discussed more thoroughly in Sec.~\ref{sec:compare} below.

\subsection{Surface charge in the absence of applied voltage}
In the absence of an applied voltage ($\bar{V}=0$), surface charges on the leaflets can still modify the effective mechanical moduli via screened electrostatic interactions:
\begin{equation}
\begin{aligned} \label{eq:correctionsSigma}
&\Lambda^{\text{eff}}=\Lambda\underbrace{+\frac{\left(\bar{\delta}-1\right)}{\lambda \bar{\delta}^2}\Gamma\varepsilon_\mathcal{M}\phi_{\rm T}^2\langle\bar{\sigma}\rangle^2(1+\alpha^2)}_{\Lambda^{\rm el}_{\sigma}}\,\\
\\
    &k^{\text{eff}}_{\text{b}}=k_{\text{b}} {\underbrace{+\frac{\lambda(2\bar{\delta}^2+4\bar{\delta}+3)\Gamma\epsilon_\mathcal{M}\phi_{\rm T}^2\langle\bar{\sigma}\rangle^2(1+\alpha^2)}{2\bar{\delta}^2}}_{k_{{\rm b},\sigma}^{\rm el}}} \ .
\end{aligned}
\end{equation}
Equation~\eqref{eq:correctionsSigma} indicates that the sign of $\Lambda^{\rm el}_\sigma$ depends critically on the ratio of the membrane thickness to the Debye length $\bar{\delta}$. In the strong screening regime where ${\bar{\delta} > 1}$, $\Lambda^{\rm eff}$ increases due to surface charge, stabilizing the membrane against perturbations. This condition coincides with physiological ionic strengths; for example, $\bar{\delta}=4$ corresponds to a Debye length ${\lambda \sim 1}$ nm ($C_0 \sim 150$ mM). Conversely, under weak screening ($\bar{\delta} < 1$), $\Lambda^{\rm el}_\sigma$ becomes negative, lowering the effective tension and favoring long-wavelength undulations---long-range electrostatic repulsion reduces the free energy of curved configurations relative to flat ones. Such weak-screening behavior arises at low ionic strengths; for instance, ${\bar{\delta}=0.4}$ corresponds to ${\lambda \sim 10\text{ nm}}$ (${C_0 \sim 1\text{ mM}}$), consistent with experimental conditions in Ref.~\cite{Riske2005}. In contrast to the tension correction, the surface-charge contribution to the bending rigidity, $k_{{\rm b},\sigma}^{\rm el}$, remains strictly positive for all $\bar{\delta}$.  The asymmetry parameter $\alpha$ amplifies the magnitude of the corrections via $(1+\alpha^2)$ but does not affect their sign.

Thus, electrostatic contributions from voltage and surface charge effects separately always stiffen bending fluctuations, even in regimes where they soften the effective tension. The tension crossover at $\bar{\delta}=1$ therefore reflects the importance of finite membrane thickness: theories that treat the membrane as a zero-thickness interface (formally $\bar{\delta}\to0$) capture only the destabilizing regime, whereas accounting for finite thickness reveals the stabilizing behavior that emerges under physiological screening conditions. 

The surface-charge corrections in the effective tension can be very large, while remaining comparatively modest for the bending rigidity. For instance, in the strong-screening regime with ${\langle\bar{\sigma}\rangle = -3.6}$ (about $8\%$ negatively charged lipids per leaflet on average), at ${\bar{\delta}=4}$, the effective tension increases by ${\sim 27,000-34,000\,\%}$ for either symmetric (${\alpha=0}$) or slightly asymmetric leaflets (${\alpha=0.5}$), whereas bending rigidity increases by a modest $11-14 \%$ under the same conditions (Fig.~\ref{fig:barplot}). In the opposite, weak-screening limit (${\bar{\delta}=0.4}$) within the same charge density regime (${\langle\bar{\sigma}\rangle = -3.6}$, ${\alpha=0}$), the tension decreases by ${54,000 \, \%}$ (Fig.~\ref{fig:barplot}), illustrating how weak screening can promote undulatory instabilities.

\subsection{Combined effects of applied voltage and surface charge}
When both surface charges and external voltage are present, the effective moduli include additional cross-terms arising from the quadratic dependence of the Maxwell stress on the electric field:
\begin{align} 
   &\Lambda^{\text{eff}}= \Lambda+\Lambda^{\rm el}_V+\Lambda^{\rm el}_{\sigma}\underbrace{-\frac{4\bar{V} \alpha \langle\bar{\sigma}\rangle(\bar{\delta}+2)\varepsilon_\mathcal{M}\phi_{\rm T}^2}{\lambda \bar{\delta}^2}}_{\Lambda^{\rm el}_{V\sigma}} \ , \nonumber\\
   \label{eq:full_corrections}\\
   &k^{\text{eff}}_{\rm b}= k_{\rm b} +k_{{\rm b},V}^{\rm el}+k_{{\rm b},\sigma}^{\rm el}\underbrace{-\frac{\lambda \bar{V} \alpha\langle\bar{\sigma}\rangle\varepsilon_{\mathcal{M}}\phi_{\rm T}^2}{\bar{\delta}^2}(3\bar{\delta}^2+8\bar{\delta}+6)}_{k_{{\rm b},V\sigma}^{\rm el}} \ . \nonumber
\end{align}
The cross-terms are proportional to $\bar{V}\alpha\langle\bar{\sigma}\rangle$ arising primarily from the charge asymmetry $\alpha$, allowing them to either enhance or offset the destabilizing voltage contribution.

The magnitude of the contributions of the cross-terms to $\Lambda^{\rm eff}$ and $k_{\rm b}^{\rm eff}$ are indicated in Fig.~\ref{fig:barplot}. Across all cases, the dominant modulation of $\Lambda^{\rm eff}$ arises from surface charge, with the voltage and mixed terms providing smaller corrections. In contrast, the bending modulus is primarily stiffened by the charge-driven term. Because instability in Eq.~(\ref{dispersion}) is determined by the small-$q$ limit, these results confirm that the system is governed by tension effects, and bending stiffening never generates an instability.

\begin{figure*}[t!]
\centering
\includegraphics[width=0.85\linewidth]{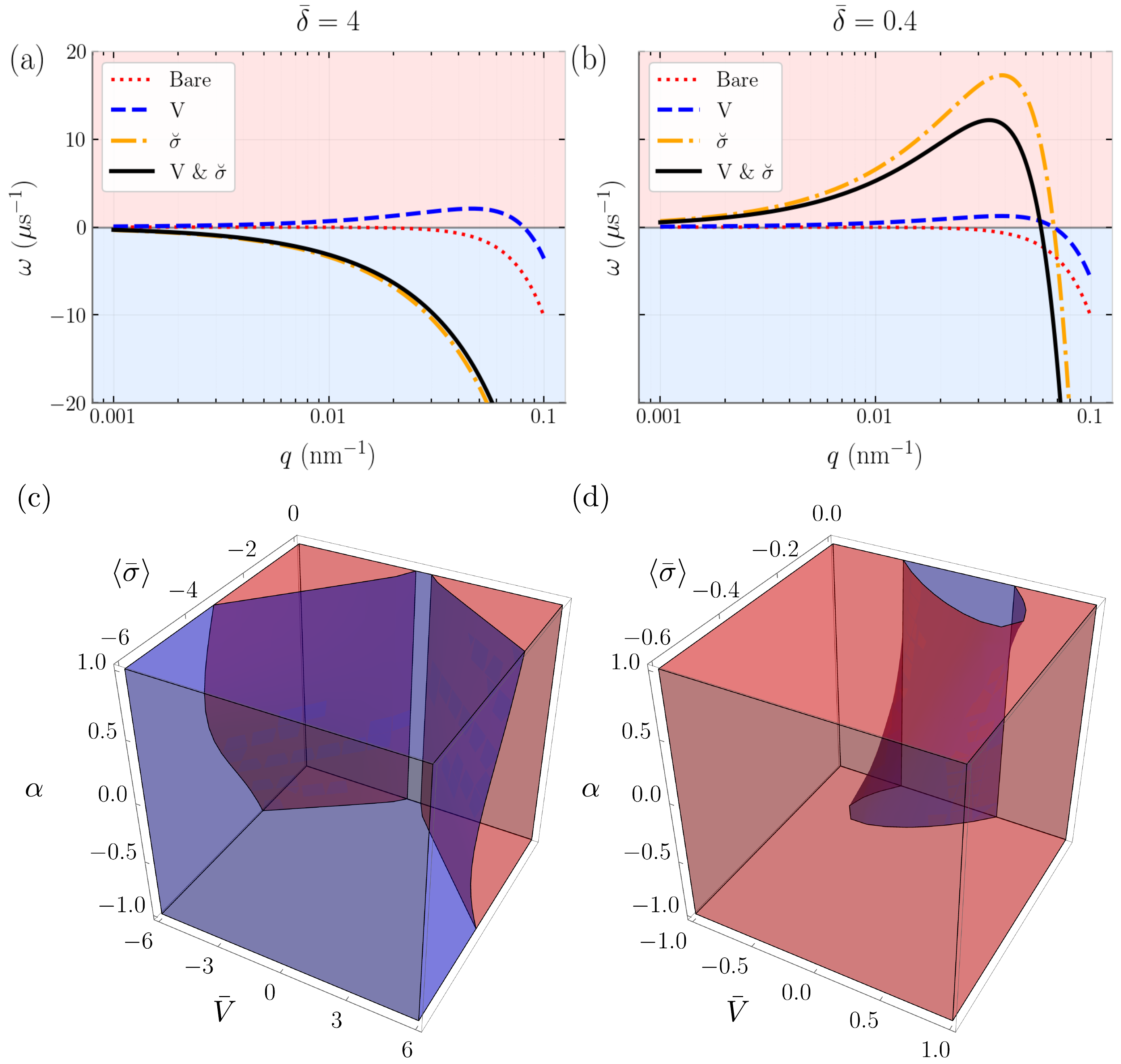}
\caption{(a,b) Dispersion relations $\omega(q)$ for four representative cases: Bare (no applied voltage and no surface charge), $\text{V}$ (nonzero voltage but no surface charge), $\breve{\sigma}$ (nonzero surface charge but no voltage), and $V \,\& \, \breve{\sigma}$ (both present). Parameter values are $\bar{V}\in \{0,3\}$, $\langle \bar{\sigma} \rangle\in \{0,-3.6\}$, and $\alpha\in \{0,0.5\}$. Panel (a) corresponds to strong screening ($\bar{\delta}=4$, $\lambda=1\,\mathrm{nm}$), and panel (b) to weak screening ($\bar{\delta}=0.4$, $\lambda=10\,\mathrm{nm}$). Regions with ${\omega(q)<0}$ indicate stable decay of fluctuations, while ${\omega(q)>0}$ signals unstable growth. (c,d) Stability diagrams in the $(\bar{V},\langle \bar{\sigma} \rangle,\alpha)$ parameter space for the same two screening regimes. Red regions denote instability ($\Lambda^{\mathrm{eff}}<0$), and blue regions correspond to stable configurations ($\Lambda^{\mathrm{eff}}>0$). Increasing the Debye length enlarges the unstable region, indicating the destabilizing influence of longer-ranged electrostatic interactions.}
\label{fig:dispersion2}
\end{figure*}

Having quantified the renormalized moduli, we now return to the full dispersion relation (Eq.~\eqref{dispersion}) to illustrate how these corrections reshape the fluctuation spectrum. Figures~\ref{fig:dispersion2}(a),(b) display representative dispersion relations for the bare membrane as well as the voltage-only, charge-only, and combined cases. These curves make explicit how each mechanism shifts the growth spectrum $\omega(q)$ predicted by Eq.~\eqref{dispersion}. The stability behavior follows directly from the sign of $\Lambda^{\rm eff}$, which controls the small-$q$ limit. When ${\bar{\delta}>1}$ (strong screening), surface charge increases the effective tension and stabilizes long-wavelength modes. Conversely, when $\bar{\delta}<1$ (weak screening), the charge-induced reduction of $\Lambda^{\rm eff}$ promotes instability. 

To summarize these trends globally, Figs.~\ref{fig:dispersion2}(c),(d) present stability diagrams in the parameter space of $\bar{V}, \langle \bar{\sigma} \rangle$, and $\alpha$ for strong ($\bar{\delta}=4$) and weak ($\bar{\delta}=0.4$) screening. Red regions denote unstable configurations with ${\Lambda^{\rm eff} < 0}$, while blue regions correspond to stable states with ${\Lambda^{\rm eff} > 0}$. As the Debye length increases, the unstable region expands, reflecting the enhanced range of electrostatic interactions.

\subsection{Validity of the Quasi-Static Approximation}
Our analysis assumes that bulk charge transport remains quasi-static. This requires the membrane charging time scale ($\tau_{\rm C}$) to be shorter than the instability growth time scale ($\tau_{\rm growth}$). The charging time corresponds to the capacitive relaxation time identified in Refs.~\cite{farhadi2025capacitive,zhao2025diffuse}. In the long-wavelength limit ($q\lambda \ll 1$), this time scale is approximately ${\tau_{\rm C} \approx \lambda /D q \bar{v}}$ (see SM Appendix C).

For weak screening (${\lambda = 10\,\mathrm{nm}}$), the dispersion relation (Fig.~\ref{fig:dispersion2}(b)) yields a maximal growth rate ${\omega_{\max} \approx 15\,\mu\mathrm{s}^{-1}}$, corresponding to ${\tau_{\rm growth} \approx 67\,\mathrm{ns}}$ at ${q \approx 0.06\,\mathrm{nm}^{-1}}$. For these parameters, one finds that the charging time is slightly shorter but of the same order, so the quasi-static approximation remains marginally valid. By contrast, under physiological conditions ($\lambda = 1\,\mathrm{nm}$; Fig.~\ref{fig:dispersion2}(a)), the instability grows more slowly, with ${\omega_{\max} \approx 2.5\,\mu\mathrm{s}^{-1}}$ (${\tau_{\rm growth} \approx 400\,\mathrm{ns}}$), while the charging time decreases to ${\tau_{\rm C} \approx 0.4\,\mathrm{ns}}$. This produces a clear separation of time scales.

Taken together, these calculations indicate that the membrane charges faster than the instability develops across all regimes considered, and the quasi-static approximation becomes increasingly accurate at physiologically relevant higher ionic strengths.

\renewcommand{\arraystretch}{3}
\begin{table*}[!t]
\centering
\small
\caption{Decomposition of the electrostatic corrections to the surface tension ($\Lambda^{\mathrm{el}}$) and bending rigidity $(k_\mathrm{b}^{\mathrm{el}}$). Predictions from prior theoretical works (rewritten in this paper's notation) are compared with the individual contributions from Term~2, corresponding to the net Maxwell traction at the membrane surfaces, and Term~1, the finite-thickness traction-moment contribution retained in the $(2+\delta)$ theory.}
\label{tab:combined_el}
\vspace{0.5em}
\textbf{(a) $\Lambda^{\mathrm{el}}$}

\vspace{0.5em}

\resizebox{\textwidth}{!}{%
{\everymath{\displaystyle}
\begin{tabular}{|l|c|c|c|}
\hline
 & \textsf{Prior Theory} & \textsf{Term 2} & \textsf{Term 1} \\
\hline
$\Lambda^{\rm el}_{\sigma}$
 & $
\begin{array}{l}
-\frac{\Gamma\varepsilon_{\mathcal{M}}\phi_{\rm T}^2}{\bar{\delta}^2\lambda}\,\langle\bar{\sigma}\rangle^{2}\!\left[1+\alpha^{2}\frac{\Gamma\bar{\delta}(\Gamma\bar{\delta}+4)}{(\Gamma\bar{\delta}+2)^{2}}\right] \text{\cite{loubet2013electromechanics}},\\[2pt]
-\frac{\Gamma\varepsilon_{\mathcal{M}}\phi_{\rm T}^2}{\bar{\delta}^2\lambda}(1+\alpha^{2})\,\langle\bar{\sigma}\rangle^{2} \text{\cite{winterhalter1988effect}}
\end{array}$
 & $\begin{array}{l}
-\frac{\Gamma\varepsilon_{\mathcal{M}}\phi_{\rm T}^2}{\bar{\delta}^2\lambda}\,\langle\bar{\sigma}\rangle^{2}\!\left[1+\alpha^{2}\frac{\Gamma\bar{\delta}(\Gamma\bar{\delta}+4)}{(\Gamma\bar{\delta}+2)^{2}}\right]\\[2pt]
\approx -\frac{\Gamma\varepsilon_{\mathcal{M}}\phi_{\rm T}^2}{\bar{\delta}^2\lambda}(1+\alpha^{2})\,\langle\bar{\sigma}\rangle^{2}\quad(\Gamma\ \text{large})
\end{array}$
 & $\begin{array}{l}
\frac{\Gamma\varepsilon_{\mathcal{M}}\phi_{\rm T}^2}{\bar{\delta}\lambda}\langle\bar{\sigma}\rangle^{2}\!\left[1+\alpha^{2}\Gamma\frac{\Gamma\bar{\delta}^{2}-2}{(\Gamma\bar{\delta}+2)^{2}}\right]\\
\approx \frac{\Gamma\varepsilon_{\mathcal{M}}\phi_{\rm T}^2}{\bar{\delta}\lambda}(1+\alpha^{2})\,\langle\bar{\sigma}\rangle^{2}\quad(\Gamma\ \text{large})
\end{array}$ \\
\hline
$\Lambda^{\rm el}_{V}$
 & $- \frac{4 \bar{V}^2 \phi_{\rm T}^2 \Gamma\varepsilon_\mathcal{M} (\Gamma\bar{\delta}+1)}{\lambda (\Gamma \bar{\delta} + 2)^2}$~\cite{yu2025instabilityfluctuatingbiomimeticmembrane, Ziebert2010_ZeroThickness, loubet2013electromechanics, ambjornsson2007applying}
 & $- \frac{4 \bar{V}^2 \phi_{\rm T}^2 \Gamma\varepsilon_\mathcal{M} (\Gamma\bar{\delta}+1)}{\lambda (\Gamma \bar{\delta} + 2)^2}$
 & $- \frac{2 \bar{V}^2 \phi_{\rm T}^2 \Gamma\varepsilon_\mathcal{M} (-2 + \Gamma) \bar{\delta}}{\lambda(\Gamma \bar{\delta} + 2)^2}$ \\
\hline
$\Lambda^{\rm el}_{V\sigma}$
 & $-\frac{4\bar{V}\varepsilon_\mathcal{M}\phi_{\rm T}^2\alpha \langle\bar{\sigma}\rangle}{\lambda \bar{\delta}^2}$~\cite{loubet2013electromechanics}
 & $
-\frac{4\bar{V}\varepsilon_\mathcal{M}\phi_{\rm T}^2\alpha \langle\bar{\sigma}\rangle}{\lambda \bar{\delta}^2}\quad(\Gamma\ \text{large})
$
 & $
-\frac{4\bar{V}(\bar{\delta}+1)\varepsilon_\mathcal{M}\phi_{\rm T}^2\alpha \langle\bar{\sigma}\rangle}{\lambda \bar{\delta}^2}\quad(\Gamma\ \text{large})
$\\
\hline
\end{tabular}}
}

\vspace{2.5em}

\textbf{(b) $k_\mathrm{b}^{\mathrm{el}}$}

\vspace{0.5em}

\resizebox{\textwidth}{!}{%
{\everymath{\displaystyle}
\begin{tabular}{|l|c|c|c|}
\hline
 & \textsf{Prior Theory} & \textsf{Term 2} & \textsf{Term 1} \\
\hline
$k^{\rm el}_{{\rm b},\sigma}$
 & $
\begin{array}{l}
\frac{\lambda(2\bar{\delta}+3)\Gamma\epsilon_\mathcal{M}\phi_{\rm T}^2\langle\bar{\sigma}\rangle^2(1+\alpha^2)}{2\bar{\delta}^2} \text{\cite{loubet2013electromechanics}},\\[2pt]
\frac{3\lambda\Gamma\epsilon_\mathcal{M}\phi_{\rm T}^2\langle\bar{\sigma}\rangle^2(1+\alpha^2)}{2\bar{\delta}^2} \text{\cite{winterhalter1988effect}}
\end{array}$
 & $\frac{\lambda(2\bar{\delta}+3)\Gamma\epsilon_\mathcal{M}\phi_{\rm T}^2\langle\bar{\sigma}\rangle^2(1+\alpha^2)}{2\bar{\delta}^2}$
 & $\frac{\lambda(\bar{\delta}+1)\Gamma\epsilon_\mathcal{M}\phi_{\rm T}^2\langle\bar{\sigma}\rangle^2(1+\alpha^2)}{\bar{\delta}}$ \\
\hline
$k^{\rm el}_{{\rm b},V}$
 & $\begin{array}{l}\frac{\bar{V}^2 \phi_{\rm T}^2 \Gamma \varepsilon_\mathcal{M} \lambda (\frac{4}{3}\Gamma \bar{\delta}^3 + 8\bar{ \delta}  + 12)}{2 (\Gamma \bar{\delta} + 2)^2}~\cite{yu2025instabilityfluctuatingbiomimeticmembrane,Ziebert2010_ZeroThickness}\\
\frac{\bar{V}^2 \phi_{\rm T}^2 \Gamma \varepsilon_\mathcal{M} \lambda (\frac{4}{3}\Gamma \bar{\delta}^3 + 8\bar{ \delta}^2 + 16\bar{\delta} + 12)}{2 (\Gamma \bar{\delta} + 2)^2}~\cite{loubet2013electromechanics,ambjornsson2007applying}
\end{array}$
 & $\frac{\bar{V}^2 \phi_{\rm T}^2 \Gamma \varepsilon_\mathcal{M} \lambda (\Gamma \bar{\delta}^3 + 8\bar{ \delta}^2 + 16\bar{\delta} + 12)}{2 (\Gamma \bar{\delta} + 2)^2}$
 & $0$ \\
\hline
$k^{\rm el}_{{\rm b}, V\sigma}$
 & $-\frac{\lambda \bar{V}\varepsilon_{\mathcal{M}}\phi_{\rm T}^2}{\bar{\delta}^2}(\frac{2}{3}\bar{\delta}^2+6\bar{\delta}+6)\alpha\langle\bar{\sigma}\rangle $~\cite{loubet2013electromechanics}
 & $
-\frac{\lambda \bar{V}\varepsilon_{\mathcal{M}}\phi_{\rm T}^2}{\bar{\delta}^2}(\bar{\delta}^2+6\bar{\delta}+6)\alpha\langle\bar{\sigma}\rangle 
$
 & $
-\frac{\lambda \bar{V}\varepsilon_{\mathcal{M}}\phi_{\rm T}^2}{\bar{\delta}}(2\bar{\delta}+2)\alpha\langle\bar{\sigma}\rangle
$\\
\hline
\end{tabular}}
}
\end{table*}
 
\subsection{Comparison with existing theories}
\label{sec:compare}
We now contextualize our results by comparing them with those of classical~\cite{Lippmann1875} and modern static theories ~\cite{winterhalter1988effect, ambjornsson2007applying, loubet2013electromechanics}, as well as dynamic approaches that assume a zero–thickness membrane~\cite{Ziebert2010_ZeroThickness, yu2025instabilityfluctuatingbiomimeticmembrane} (also see SM Sec. IV).

In the $(2+\delta)$-dimensional framework, the dispersion relation (Eq.~\ref{dispersion}) originates from the linearized shape equation (Eq.~\ref{eq:shape_equation}), which couples mechanical and electrostatic stresses in a fluctuating finite-thickness bilayer.  Beyond the classical Helfrich terms, the linearized shape equation contains three electromechanical contributions. However, one of these terms:
\[ -\delta H(\boldsymbol{t}^+ - \boldsymbol{t}^-) \cdot \boldsymbol{n} \ ,\]
vanishes identically for linear stability about an initially flat membrane (with base curvature ${H=0}$), and therefore does not contribute to Eq.~\eqref{dispersion} at leading order. Thus, we focus on the two remaining terms in Eq.~\eqref{eq:shape_equation} that control the fluctuation spectrum:
\[
\underbrace{Q^\alpha_{;\alpha}}_{\text{Term 1: thickness moment}} 
\quad\text{and}\quad
\underbrace{(\boldsymbol{t}^{+} + \boldsymbol{t}^{-})\cdot\boldsymbol{n}}_{\text{Term 2: surface traction}}.
\]
Tables~\ref{tab:combined_el} isolates the contributions from Term~1 and Term~2 to $\Lambda^{\rm eff}$ and $k_{\rm b}^{\rm eff}$, and place them alongside prior theoretical results to make their correspondences and differences transparent. 

Term 2 may be understood first, as it represents a net normal traction containing hydrodynamic contributions from the bulk and the Maxwell stress jump at the upper and lower membrane interfaces (cf. Eq.~\eqref{eq:shape_equation}).  In appropriate limits, this term reduces to the Lippmann electrocapillary expression $-\tfrac{1}{2} C_{\mathcal M} \Delta V_{\mathcal M}^2$~\cite{Lippmann1875} for tension corrections in thin membranes under weak screening ($\delta \to 0$). It is also consistent with the electrostatic tension and bending corrections obtained from equilibrium Poisson–Boltzmann calculations near curved charged membranes~\cite{winterhalter1988effect, winterhalter1992bending, lekkerkerker1989contribution, ziebert2010poisson}, and with finite-thickness dielectric-slab models that explicitly resolve electric fields within the membrane~\cite{ambjornsson2007applying, loubet2013electromechanics}. In dynamic settings, when the membrane thickness becomes small compared to other length scales, Term~2 also recovers the electrostatic contributions to surface tension obtained in zero-thickness electrohydrodynamic and PNP formulations~\cite{Ziebert2010_ZeroThickness, lacoste2009electrostatic, yu2025instabilityfluctuatingbiomimeticmembrane}. Thus, Term~2 captures the electrostatic force arising from field discontinuities across the membrane that is common to prior equilibrium, finite-thickness electrostatic, and zero-thickness dynamic theories, although those approaches derive it using different physical and mathematical formalisms.
\begin{figure*}[!t]
\centering
\includegraphics[width=\linewidth]{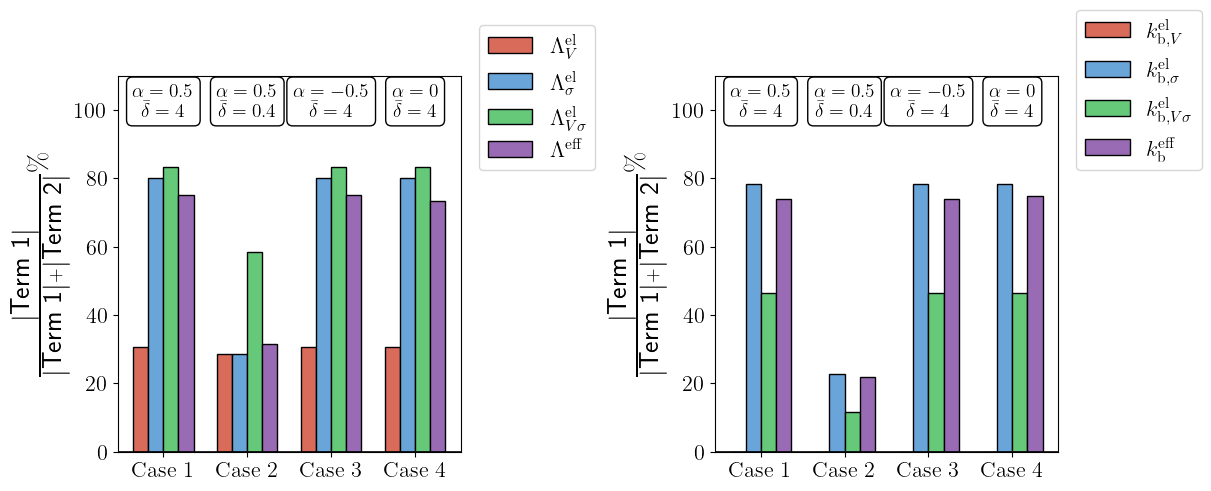}
\caption{
Bar charts showing the percentage contribution of Term~1 to the total electrostatic renormalization (Term~1 + Term~2) of the surface tension $\Lambda$ (left) and bending rigidity $k_{\rm b}$ (right) for the voltage-driven, charge-driven, and mixed ($V\sigma$) mechanisms. Results are shown for the same four membrane conditions (Cases~1–4) as in Fig.~\ref{fig:barplot}. The dimensionless voltage is fixed at $\bar{V}=3$, and the mean dimensionless surface charge density at $\langle\bar{\sigma}\rangle=-3.6$. The corresponding charge asymmetry $\alpha$ and dimensionless thickness $\bar{\delta}$ for each case are indicated in the figure annotations.}
\label{fig:term1_term2}
\end{figure*}

The essential new contribution due to the $(2+\delta)$-dimensional approach is Term~1. This term represents the divergence of a surface traction moment—an electromechanical coupling internal to the membrane that emerges from reducing the three-dimensional stress balance across a finite thickness to a mid-surface description (see SM Sec. II.5). It is evident from Eq.~\eqref{eq:def_Q_alpha} that Term~1 vanishes for a zero-thickness membrane. Physically, it acts as a transverse-shear stress resultant, defined as the first moment of the stress discontinuity between the membrane and the adjacent fluids. Because $Q^\alpha$ has dimensions of force per unit length, its divergence $Q^{\alpha}_{;\alpha}$ produces a normal force density whenever transverse shear varies spatially. Crucially, this term generates curvature-dependent stresses even in regimes where surface theories predict none. These stresses renormalize both the surface tension and bending rigidity, producing quantitative differences relative to prior models, summarized in 
\km{Table~\ref{tab:combined_el}.}

To quantify the numerical importance of this correction, Fig.~\ref{fig:term1_term2} shows the percentage contribution of Term 1 to the total electrostatic correction (Term 1 + Term 2) for the same four membrane conditions considered in Fig.~\ref{fig:barplot}. The left and right panels in Fig.~\ref{fig:term1_term2}, which correspond to the tension and bending corrections respectively, show that contribution from Term~1 is not asymptotically small: rather, under physiological screening (${\bar{\delta}\sim4}$) it accounts for the major fraction (${>70}$ \%) of the total correction for both $\Lambda$ and $k_{\rm b}$. Under weaker screening conditions ($\bar{\delta}\sim0.4$), the relative contribution decreases but remains non-negligible, at roughly ${\sim 20}$ \% across representative cases. This confirms that finite-thickness electromechanical coupling remains quantitatively significant under both physiological and experimentally relevant conditions.

\begin{table*}[t]
\caption{Breakdown voltage comparison. Experimental $V_{\rm bd}$ values from Refs.~\cite{huang1964formation,MiyamotoThompson1967,Needham1995} alongside predictions from the $(2+\delta)$ theory and from prior surface/zero-thickness models, obtained by applying the instability condition $\Lambda^{\rm eff}=0$ in Eq.~\eqref{eq:full_corrections}.}
\label{tab:breakdown}
\begin{ruledtabular}
\begin{tabular}{lcccccccc}
Lipid System &$\Lambda$(pN/nm)\footnotemark[1]&$\bar{\delta}$&$\lambda$(nm)&$\langle\bar{\sigma}\rangle$&$\alpha$& $V_{\rm bd}$(mV) (measured) & $V_{\rm bd}$(mV) (predicted)\footnotemark[2] &$V_{\rm bd}$(mV) (predicted)\footnotemark[3]\\
\hline
Egg Phosphatidyl &0.5 &3.7-6.1&1&0&0& 200~\cite{huang1964formation, MiyamotoThompson1967}  & 233--298&283-362\\
SOPC:Cholesterol &11-15.4&2.47& 3&-3&0& 1950~\cite{Needham1995} & 1740--2051 &2058-2442\\
RBC lipids  &8.9-10.5  &2.4&  3   &-3&0& 1600~\cite{Needham1995} & 1550--1679 &1821-1982\\
SOPC        &5.5-5.9   &2.57& 3  &-3&0& 1100~\cite{Needham1995} & 1271--1314 &1470-1525\\
\end{tabular}
\end{ruledtabular}
\footnotetext[1]{Since the membrane bare tension is not provided in the text, we use the lysis tension as a substitute. }
\footnotetext[2]{This is the breakdown voltage predicted by our theory using Table \ref{tab:combined_el}. }
\footnotetext[3]{This is the breakdown voltage predicted by the prior theories using Table \ref{tab:combined_el}. }
\end{table*}

\subsection{Comparison with breakdown-voltage measurements}
\label{sec:breakdown}
To relate the linear stability analysis to experimentally measured electroporation thresholds, we estimate the breakdown voltage $V_{\rm bd}$ by setting the effective tension ${\Lambda^{\rm eff}=0}$ in Eq.~\eqref{eq:full_corrections}, which defines the onset of the long-wavelength instability. We compare these predictions with classic \emph{black lipid membrane} measurements in planar bilayers \cite{huang1964formation,MiyamotoThompson1967} and giant unilamellar vesicle experiments in which membrane tension is controlled by micropipette aspiration prior to electrical pulsing \cite{Needham1995}. Table~\ref{tab:breakdown} summarizes the measured breakdown voltages alongside predictions from the $(2+\delta)$ theory and from prior zero-thickness surface models.

For the egg phosphatidyl black lipid membrane system, reported material parameters allow direct parameterization of the theory. For the remaining lipid systems, the bare membrane tension is not explicitly reported; to enable comparison, we use the experimentally measured lysis tension at zero applied voltage (see Fig. 3 in Ref.~\cite{Needham1995}) as the representative baseline tension. In all cases, we set the charge asymmetry parameter ${\alpha=0}$ because leaflet-resolved charge asymmetry is not available in the cited datasets.

Across all lipid systems considered, the $(2+\delta)$ predictions lie closer to the measured breakdown voltages than those obtained from zero-thickness surface theories (Table~\ref{tab:breakdown}). The improved agreement arises from the finite-thickness traction-moment contribution (Term~1). This term provides the dominant correction to the voltage-dependent renormalization of the effective tension when the membrane thickness $\delta$ is comparable to the Debye length $\lambda$ (Fig.~\ref{fig:term1_term2}). As a result, accounting for finite-thickness effects shifts the predicted instability threshold by 10-30 \%, improving the agreement with experiment.

\subsection{Density fluctuations}
The preceding analysis considered undulatory instabilities in an incompressible membrane, where the dynamics is governed solely by height fluctuations. However, such height fluctuations may lead to local variations in lipid packing, generating density fluctuations that modify membrane stress and may result in pore formation. To study the coupling between height and density fluctuations, we relax the assumption of strict area incompressibility and extend the surface balance laws to include the areal mass density $\rho_\mathrm{s}$ as a dynamical variable.  For small perturbations about a flat, constant density base state, i.e., ${\rho_\mathrm{s} = \breve{\rho}_\mathrm{s} + \epsilon \tilde{\rho}_\mathrm{s}}$, where $\breve{\rho}_\mathrm{s}$ is the base state density and $\tilde{\rho}_\mathrm{s}$ is the perturbation variable. At leading order, surface mass conservation satisfies:
\begin{equation}
\partial_t\tilde{\rho}_\mathrm{s}+\breve{\rho}_\mathrm{s}\,\tilde v^{\alpha}_{0;\alpha}=0 \ . \label{perturbed_massbalance}   
\end{equation}
Upon linearization and normal mode decomposition, each in-plane mode with wavevector $\boldsymbol{q}$ obeys a coupled set of equations for the height and density perturbations. The resulting dynamics contains two characteristic relaxation rates, $\alpha_1(q)$ and $\omega(q)$:
\begin{equation}
\begin{aligned}
&h_{\mathbf{q}}(t) = h_{\mathbf{q0}}\, e^{\omega(q) t} \ , \\
&\tilde{\rho}_{\mathrm{s}\mathbf{q}}(t) = C_1 e^{\alpha_1(q) t} + \frac{\mathcal{F}(q)\, h_{\mathbf{q0}}e^{\omega(q) t}}{(\alpha_1(q) - \omega(q))} \ ,\label{density solution}
\end{aligned}
\end{equation}
where $\omega(q)$ is specified in Eq.~\eqref{dispersion}, $C_1$ is a constant determined by the initial condition, and
\begin{equation}
\alpha_1(q) = -\frac{2 \bar{k}_{\rm c} q}{4\mu + q(\bar{w} + 2\zeta)}\label{density fluctuation}
\end{equation}
and in the long wavelength limit (${q\lambda \ll 1}$):
\begin{equation}
    \mathcal{F}(q)=\frac{4\breve{\rho}_\mathrm{s}\Gamma\varepsilon_\mathcal{M}\phi_{\rm T}^2 \langle\bar{\sigma}\rangle (-2\bar{V}+\alpha\langle\bar{\sigma}\rangle\Gamma)q}{(\Gamma\bar{\delta}+2)(4\mu+q(2\zeta+\bar{w}))\lambda\delta} \label{F term} 
\end{equation}
Here $\bar{k}_{\rm c}$ is the area compression modulus, $\zeta$ and $\bar{w}$ are the membrane shear and dilatational viscosities, respectively, and $\mathcal{F}(q)$ is an electrostatic forcing coefficient proportional to the product of applied voltage and charge asymmetry (see SM Sec. V for the full derivation). Importantly, the relaxation rate $\alpha_1(q)$ depends only on mechanical parameters and viscous dissipation, and is therefore unaffected by electrostatic renormalization. Physically, $\alpha_1$ sets how quickly lipids rearrange in-plane to accommodate local area changes, whereas $\omega$ dictates the relaxation of the underlying height mode that drives density changes. To leading order, the height dynamics remain unchanged by density fluctuations, as indicated by Eq.~\eqref{density solution}.

Two regimes follow immediately. When the membrane is linearly stable ($\Lambda^{\mathrm{eff}} > 0$), height and density modes decay for all wavevectors. When ${\left| \alpha_1 \right| < \left| \omega \right|}$, the density and height modes decay at different timescales. By contrast, when ${\left| \alpha_1 \right| > \left| \omega \right|}$, $\tilde{\rho}_{\mathrm{s}\mathbf{q}}(t)$  tracks $h_{\mathbf{q}}(t)$ after a brief transient. Near the undulatory instability ($\Lambda^{\mathrm{eff}} \to 0^-$), density fluctuations become adiabatically attached to the growing height mode, so that the instability of the height field necessarily induces a corresponding instability in the density modes. In the unstable regime, the dominant contribution to the dynamics arises from the wavevector that maximizes the growth rate of the height mode ${q_* \approx \sqrt{-2 \Lambda^{\rm eff} / 3 k_b^{\rm eff}}}$ (saddle point approximation).

\begin{figure}
\centering
\includegraphics[width=\linewidth]{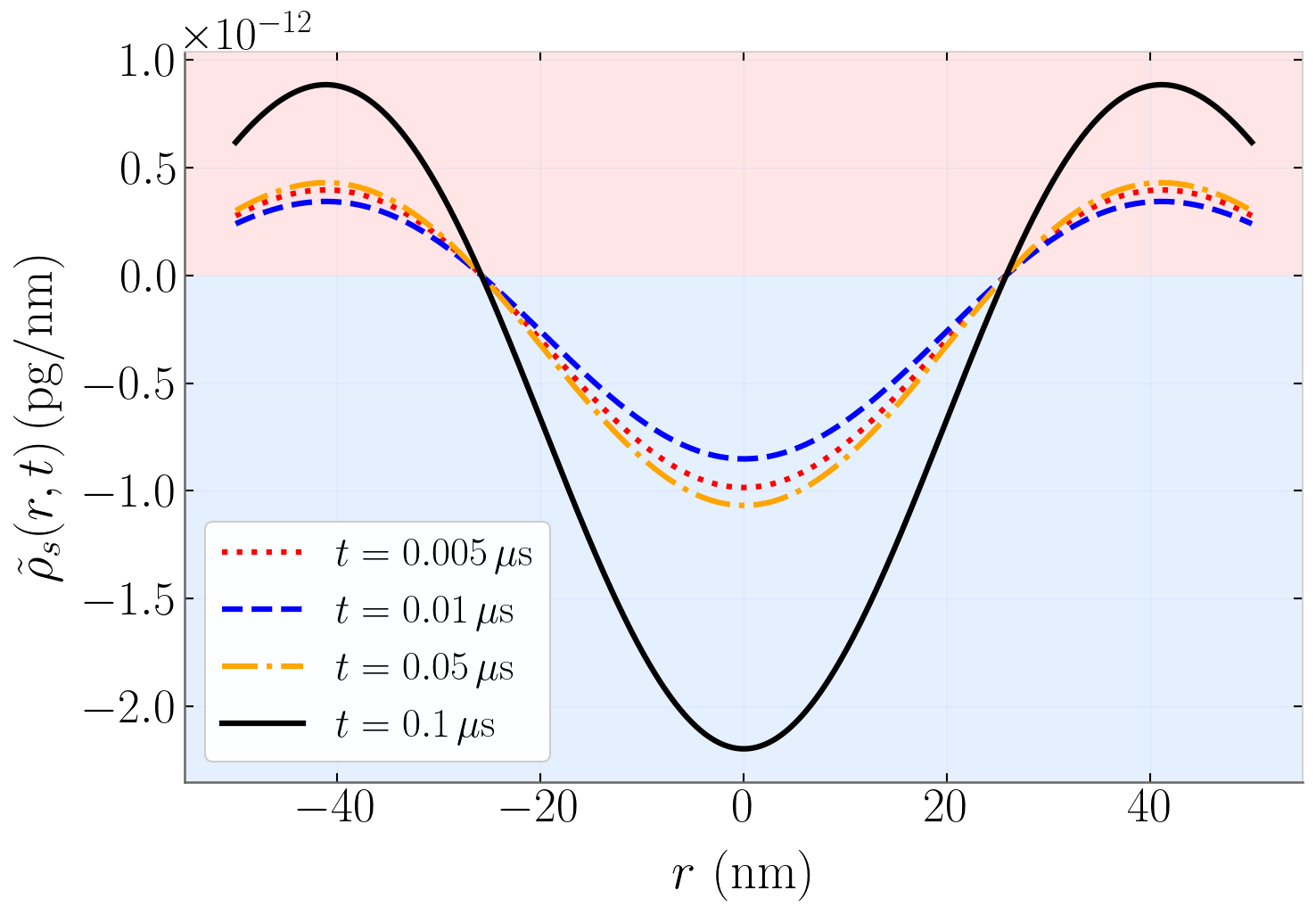}
\caption{Time evolution of density fluctuations in a biological membrane subjected to an applied voltage $\bar{V}=10$. The mean surface charge density is $\langle \bar{\sigma} \rangle = -3.6$, and asymmetric parameter is $\alpha = 0.5$. The plot is obtained from Eqs.~\eqref{density solution}, \eqref{density fluctuation}, and \eqref{F term}, setting $h_{\mathbf{q0}}=1$ (corresponding to $h(x,y,0)=\delta(x)\delta(y)$) and $C_1=0$. Curves from early to late times are shown as red dotted, blue dashed, orange dash-dotted, and black solid lines, respectively.}
\label{fig:densityplot}
\end{figure}

To illustrate the spatiotemporal evolution of density fluctuations, we consider an initially localized height perturbation, ${h(x, y, 0) = \delta(x)\delta(y)}$, and follow the resulting evolution of $\tilde{\rho}_\mathrm{s}(x,y,t)$. Using a saddle-point approximation to the inverse Fourier transform (see SM Sec. V for details), we obtain the approximate solution shown in Fig.~\ref{fig:densityplot}. The figure shows both the initial transient regime, during which density relaxes independently according to $\alpha_1(q)$, and the subsequent growth of density perturbations driven by the unstable height mode. For this case, the dominant mode occurs at ${q_*\approx 0.09 \text{ nm}^{-1}}$, with growth rates $\alpha_1(q_*)=336\,\mu\text{s}^{-1},\, \omega(q_*)=20.7\,\mu\text{s}^{-1}$.

Electrostatics influences density fluctuations in two distinct ways. First, electrostatic renormalization of $\Lambda^{\mathrm{eff}}$ and $k_{\mathrm{b}}^{\mathrm{eff}}$ modifies $\omega(q)$ and therefore the dominant wavenumber $q_*$. By contrast, $\alpha_1(q)$ remains unchanged. When the transmembrane voltage and leaflet charge asymmetry have opposite signs, increasing the voltage drives $\Lambda^{\mathrm{eff}}$ toward more negative values, shifting the unstable spectrum toward shorter wavelengths and increasing the growth rate. Second, the forcing term that couples $h_{\boldsymbol{q}}$ into $\tilde{\rho}_{\mathrm{s}\boldsymbol{q}}$ (Eq.$~\eqref{F term}$) is proportional to $\bar{V} \langle \bar{\sigma}\rangle$ and $-\langle \bar{\sigma}\rangle^2 \alpha$, reflecting the quadratic (Maxwell) nature of the electrostatic stress, and setting the amplitude of the density response. Notably, when the membrane carries no net charge (i.e., ${\langle \bar{\sigma} \rangle = 0}$), the forcing term vanishes so that ${\mathcal{F}(q)=0}$ in Eq.~\eqref{F term}. In this limit, density fluctuations decouple from height fluctuations and relax diffusively since ${\alpha_1(q) > 0}$.
Finally, the two-rate structure provides a framework for understanding pore nucleation under electrical loading without assuming a priori separation of timescales. The relative magnitudes of $\alpha_1(q)$ and $\omega(q)$ depend on membrane compressibility, viscosity, and electrostatic renormalization, and therefore determine whether density relaxes independently or remains tied to the height mode. As mentioned before, when $|\omega(q_*)| \ll |\alpha_1(q_*)|$, the dynamics is dominated by the height-driven contribution. Conversely, when $|\omega(q_*)| \gg |\alpha_1(q_*)|$, the response is governed by in-plane compressibility. 

These regimes may correspond to distinct physical routes to membrane poration. Within the $(2+\delta)$ framework, tuning $\Lambda^{\mathrm{eff}}$ toward zero increases the dominant wavelength and reduces $|\omega|$, thereby broadening the temporal window over which density fluctuations can develop and potentially promoting porogenesis. By contrast, when $|\omega(q_*)|$ becomes sufficiently large, the rapid amplification of the height mode may drive the system into a strongly nonlinear regime, potentially leading to rupture rather than pore formation. Determining the dynamical boundary between these two failure pathways remains an open question and represents an important direction for future investigation.

\section{Discussion}
\label{sec:discussion}
This article shows how applied voltages and surface charges, arising from charged lipids or ion adsorption, affect the stability of membrane fluctuations. To this end, we apply the recently developed $(2+\delta)$-dimensional theory that accounts for the finite thickness of lipid membranes, required for accurately capturing electromechanical effects. As opposed to previous theories, this framework remains valid even when the membrane thickness $\delta$ is comparable to the electrostatic screening length $\lambda$, a condition consistent with physiological ionic strengths where $\delta/\lambda = O(1)$, and permits studying the dynamical behavior of membranes. A stability analysis yields the closed-form dispersion relation in Eq.~\eqref{dispersion}, which shows that electrical loading renormalizes the surface tension and bending modulus to $(\Lambda^{\rm eff},k_b^{\rm eff})$.  
Instability is determined by the sign of $\Lambda^{\rm eff}$: when $\Lambda^{\rm eff}<0$, long-wavelength modes grow. In contrast, electrostatic effects generally produce a net increase in $k_b^{\rm eff}$, stiffening the membrane.

Beyond the familiar net Maxwell traction term $(\boldsymbol{t}^+ + \boldsymbol{t}^-) \cdot \boldsymbol{n}$ that appears in zero-thickness surface theories \cite{Ziebert2010_ZeroThickness, yu2025instabilityfluctuatingbiomimeticmembrane}, the $(2+\delta)$ dimension reduction introduces a traction-moment contribution $Q^\alpha_{;\alpha}$. This term may be interpreted as the first thickness moment of the stress discontinuity across the membrane (See SM Sec. II), and is absent from zero-thickness models. Importantly, our work suggests that under physiological conditions, the correction provided by this term is not small: for $\delta/\lambda\sim 4$, it makes the dominant contribution to the electrostatic renormalization of both $\Lambda$ and $k_\mathrm{b}$ (Fig.~\ref{fig:term1_term2}). The order-unity differences relative to PNP-based and leaky-dielectric surface theories arise primarily from the $Q^\alpha_{;\alpha}$ contribution, demonstrating that finite-thickness stress moments materially modify the effective moduli even when the membrane is geometrically thin.

The theory further reveals a screening-dependent—and therefore experimentally tunable—role of surface charge in membrane stability that is otherwise absent in prior models. The surface-charge contribution to $\Lambda^{\rm eff}$ changes sign with $\bar{\delta}=\delta/\lambda$. For $\bar{\delta}>1$ (strong screening), leaflet charge increases $\Lambda^{\rm eff}$ and stabilizes long-wavelength modes. For $\bar{\delta}<1$ (weak screening), the same charge decreases $\Lambda^{\rm eff}$ and promotes instability. Thus, increasing ionic strength screens electrostatic interactions and restores stability, whereas in ultradilute electrolytes long-range electrostatic forces can overcome elastic restoring stresses. This sign reversal follows directly from the finite-thickness stress balance and is not captured in thin-film limits that take $\bar{\delta}\to 0$. In contrast, $k_\mathrm{b}^{\rm eff}$ is increased by surface charge for all $\bar{\delta}$, reinforcing that instability remains tension-dominated. Ionic strength therefore provides a practical control parameter for shifting the stability boundary at fixed charge density.

When applied voltage and fixed charge coexist, mixed contributions scale as $\bar{V}\alpha\langle\bar{\sigma}\rangle$, introducing a directional dependence on leaflet asymmetry. This dependence is enabled by the finite-thickness formulation: because the two membrane interfaces are retained explicitly, the theory can accommodate distinct surface charge densities on the two leaflets. By contrast, conventional zero-thickness surface models collapse the bilayer into a single interface and therefore cannot naturally represent leaflet-resolved surface electrostatics. The sign of $\alpha$ shifts the instability threshold in a manner absent from symmetric models, providing a direct route by which leaflet-resolved electrochemistry modifies breakdown conditions. Near the instability boundary, the coupled dynamics of height and density indicate that areal density fluctuations become slaved to growing undulatory modes, selecting a lateral thinning scale that may act as a precursor to pore nucleation. In this sense, the linear instability condition $\Lambda^{\rm eff}=0$ offers a mechanistic bridge between field-induced membrane softening and experimentally observed electroporation timescales, without invoking an independent nucleation hypothesis.

We close by noting a few caveats. The present analysis is carried out about a planar base state. This assumption is appropriate for the parameter regimes examined here: prior studies~\cite{PhysRevE.64.051922} indicate that spontaneous curvature-driven vesiculation arises only at surface charge densities substantially larger than those considered in this work. Applying the criterion of Ref.~\cite{PhysRevE.64.051922}, the threshold for spherical instability corresponds to $\langle\bar{\sigma}\rangle \approx 7$ for $\lambda=10,\mathrm{nm}$ and $\langle\bar{\sigma}\rangle \approx 70$ for $\lambda=1,\mathrm{nm}$—both far exceeding physiologically relevant charge densities. Within the experimentally accessible range studied here, the flat configuration therefore remains the appropriate reference state. At sufficiently high charge densities beyond this regime, however, curvature selection and vesiculation would occur prior to the undulatory instability analyzed here, pointing toward a nonlinear extension of the present framework.

We also adopt a quasi-static approximation for diffuse-layer charging. As shown in Sec.~\ref{sec:results}, capacitive charging proceeds on timescales comparable to or faster than the growth of unstable modes, and under physiological ionic strengths the separation of timescales is particularly strong. Although the present work is restricted to linear stability and does not address nonlinear pore growth or edge energetics, the instability thresholds and renormalized moduli derived here provide a quantitative basis for future studies of fully nonlinear electromechanical evolution.

A direct experimental test of these predictions can be realized by adapting the freestanding lipid bilayer tensiometer recently designed by P\'erez-Mitta and MacKinnon~\cite{doi:10.1073/pnas.2221541120}. In this setup, membrane tension can be extracted from curvature via the Young–Laplace relation while capacitance measurements are performed under applied voltage. Because the instability criterion derived here is $\Lambda^{\rm eff}=0$, this setup enables direct tension-resolved tracking of electrostatic softening prior to pore nucleation. By independently varying leaflet charge density (through lipid composition) and Debye screening length (through ionic strength), one could construct a stability diagram in $(\bar V,\langle\bar{\sigma}\rangle,\lambda)$ space and directly probe the finite-thickness traction-moment contribution---the unique signature of the $(2+\delta)$ framework---which is predicted to dominate when $\delta/\lambda=O(1)$. Such measurements would provide a quantitative bridge between electromechanical fluctuation spectra and experimentally observed breakdown thresholds.

\section*{Acknowledgments}

The authors are grateful to Joshua Fernandes, Jafar Farhadi and Dr. Hyeongjoo Row for helpful discussions. All results were originally derived by the authors. Large language models (ChatGPT and Claude) were used to improve textual clarity in some parts of the manuscript and to assist in checking intermediate steps in some analytical calculations. Symbolic and numerical calculations were performed using Mathematica and Python. All scientific content, derivations, and conclusions were independently verified by the authors. S.N ackowledges support from the Gordon and Betty Moore Foundation.
K.K.M is supported by Director, Office of Science, Office of Basic Energy Sciences, of the U.S. Department of Energy under contract No. DEAC02-05CH11231. K.S. acknowledges support from the Hellman Foundation, the McKnight Foundation, the Alfred P. Sloan Foundation, and the University of California, Berkeley.

\bibliography{reference}

\end{document}